\documentclass[a4paper,fleqn,usenatbib]{mnras}

\usepackage[T1]{fontenc}
\usepackage{ae,aecompl}
\usepackage{graphicx}
\usepackage{amsmath}
\usepackage{amssymb}

\newcommand{\Msun}{\ensuremath{\,{M}_\odot}}                            
\newcommand{\Rsun}{\ensuremath{\,{R}_\odot}}                            
\newcommand{\Mjup}{\ensuremath{\,{M}_{\rm Jup}}}                        
\newcommand{\Rjup}{\ensuremath{\,{R}_{\rm Jup}}}                        
\newcommand{\Teq}{\ensuremath{T_{\rm eq}^{\,\prime}}}                   
\newcommand{\safronov}{\ensuremath{\Theta}}                             
\newcommand{\mss}{\,m\,s$^{-2}$}                                        
\newcommand{\pjup}{\ensuremath{\,\rho_{\rm Jup}}}                       
\newcommand{\psun}{\ensuremath{\,\rho_\odot}}                           

\title[star-spots on WASP-52]{Orbital alignment and star-spot properties in the WASP-52 planetary system}
\author[L. Mancini et al.]{
\hspace{-0.34cm}
L.\ Mancini$^{1,\,2}$\thanks{E-mail: mancini@mpia.de},
J.\ Southworth$^{3}$,
G.\ Raia$^{4}$,
J.\ Tregloan-Reed$^{5}$,
P.\ Molli\`{e}re$^{1}$,
V.\ Bozza$^{6,7}$,
\newauthor
M.\ Bretton$^{8}$,
I.\ Bruni$^{9}$,
S.\ Ciceri$^{1}$,
G.\ D'Ago$^{6,7,10}$,
M.\ Dominik$^{11}$,
T.\ C.\ Hinse$^{12}$,
\newauthor
M.\ Hundertmark$^{13}$,
U.\ G.\ J{\o}rgensen$^{13}$,
H.\ Korhonen$^{14,13,15}$,
M.\ Rabus$^{16,1}$,
S.\ Rahvar$^{17}$,
\newauthor
D.\ Starkey$^{11}$,
S.\ {Calchi Novati}$^{6,18,10}$,
R.\ {Figuera Jaimes}$^{11,19}$,
Th.\ Henning$^{1}$,
D.\ Juncher$^{13}$,
\newauthor
T.\ Haugb{\o}lle$^{13}$,
N.\ Kains$^{20}$,
A.\ Popovas$^{13}$,
R.\ W.\ Schmidt$^{21}$,
J.\ Skottfelt$^{22, 13}$,
\newauthor
C.\ Snodgrass$^{23}$,
J.\ Surdej$^{24}$,
O.\ Wertz$^{24}$ \\
\\
$^{1}$Max Planck Institute for Astronomy, K\"{o}nigstuhl 17, D-69117 Heidelberg, Germany, E-mail: mancini@mpia.de \\
$^{2}$INAF -- Osservatorio Astrofisico di Torino, via Osservatorio 20, I-10025 Pino Torinese, Italy \\
$^{3}$Astrophysics Group, Keele University, Keele ST5 5BG, UK \\
$^{4}$INAF -- Osservatorio Astronomico di Capodimonte, via Moiariello 16, I-80131 Naples, Italy \\
$^{5}$NASA Ames Research Center, Moffett Field, CA 94035, USA \\
$^{6}$Dipartimenti di Fisica ``E.\,R. Caianiello'', Universit\`a di Salerno, Via Giovanni Paolo II 132, I-84084 Fisciano (SA), Italy \\
$^{7}$Istituto Nazionale di Fisica Nucleare, Sezione di Napoli, Via Cintia, I-80126 Naples, Italy \\
$^{8}$Observatoire des Baronnies Proven\c{c}ales, Le Mas des Gr\`{e}s, Route de Nyons, F-05150 Moydans, France\\
$^{9}$INAF -- Osservatorio Astronomico di Bologna, Via Ranzani 1, I-40127 Bologna, Italy \\
$^{10}$International Institute for Advanced Scientific Studies (IIASS), Via G.\ Pellegrino 19, I-84019 Vietri sul Mare (SA), Italy \\
$^{11}$SUPA, School of Physics \& Astronomy, University of St Andrews, North Haugh, St Andrews, Fife KY16 9SS, UK \\
$^{12}$Korea Astronomy \& Space Science Institute, 776 Daedukdae-ro, Yuseong-gu, 305-348 Daejeon, Republic of Korea \\
$^{13}$Niels Bohr Institute \& Centre for Star and Planet Formation, University of Copenhagen, {\O}ster Voldgade 5, DK-1350 Copenhagen, Denmark \\
$^{14}$Dark Cosmology Centre, Niels Bohr Institute, University of Copenhagen, Juliane Maries Vej 30, DK-2100 Copenhagen, Denmark \\
$^{15}$Finnish Centre for Astronomy with ESO (FINCA), V{\"a}is{\"a}l{\"a}ntie 20, FI-21500 Piikki{\"o}, Finland \\
$^{16}$Instituto de Astrof\'{i}sica, Pontificia Universidad Cat\'{o}lica de Chile, Av. Vicu\~{n}a Mackenna 4860, 7820436 Macul, Santiago, Chile \\
$^{17}$Department of Physics, Sharif University of Technology, PO Box 11155-9161 Tehran, Iran \\
$^{18}$NASA Exoplanet Science Institute, MS 100-22, California Institute of Technology, Pasadena, CA 91125, USA \\
$^{19}$European Southern Observatory, Karl-Schwarzschild-Strasse 2, D-85748 Garching bei M\"{u}nchen, Germany \\
$^{20}$Space Telescope Science Institute, 3700 San Martin Drive, Baltimore, MD 21218, USA \\
$^{21}$Astronomisches Rechen-Institut, Zentrum f\"{u}r Astronomie, Universit\"{a}t Heidelberg, M\"{o}nchhofstrasse 12-14, D-69120 Heidelberg, Germany \\
$^{22}$Centre for Electronic Imaging, Department of Physical Sciences, The Open University, Milton Keynes MK7 6AA, UK \\
$^{23}$Planetary and Space Science, Department of Physical Sciences, The Open University, Milton Keynes MK7 6AA, UK \\
$^{24}$Institut d'Astrophysique et de G\'eophysique, All\'ee du 6 Ao\^ut 19c, Sart Tilman, B-4000 Li\`ege, Belgium \\
}

\date{Accepted XXX. Received YYY; in original form ZZZ}

\pubyear{2016}

\begin{document}
\label{firstpage}
\pagerange{\pageref{firstpage}--\pageref{lastpage}}
\maketitle

\begin{abstract}
We report 13 high-precision light curves of eight transits of the exoplanet WASP-52 b, obtained by using four medium-class telescopes, through different filters, and adopting the defocussing technique. One transit was recorded simultaneously from two different observatories and another one from the same site but with two different instruments, including a multi-band camera. Anomalies were clearly detected in five light curves and modelled as star-spots occulted by the planet during the transit events. We fitted the clean light curves with the {\sc jktebop} code, and those with the anomalies with the {\sc prism+gemc} codes in order to simultaneously model the photometric parameters of the transits and the position, size and contrast of each star-spot. We used these new light curves and some from the literature to revise the physical properties of the WASP-52 system. Star-spots with similar characteristics were detected in four transits over a period of 43\,d. In the hypothesis that we are dealing with the same star-spot, periodically occulted by the transiting planet, we estimated the projected orbital obliquity of WASP-52\,b to be $\lambda=3^{\circ}.8 \pm 8^{\circ}.4$. We also determined the true orbital obliquity, $\psi=20^{\circ} \pm 50^{\circ}$, which is, although very uncertain, the first measurement of $\psi$ purely from star-spot crossings. We finally assembled an optical transmission spectrum of the planet and searched for variations of its radius as a function of wavelength. Our analysis suggests a flat transmission spectrum within the experimental uncertainties.

\end{abstract}



\section{Introduction}
\label{sect_01}

Among all extrasolar planets, those that are transiting are recognised as the most interesting to study in detail. The fact that they periodically transit their parent stars makes it possible to measure their physical and orbital parameters with exquisite precision by means of standard astronomical techniques. These measurements can also include their spin-orbit alignment (i.e. the sky-projected angle between planetary orbital axis and stellar spin, $\lambda$), their thermal flux and reflected light, and the chemical composition of their atmosphere. These parameters are precious for theoretical astrophysicists seeking to understand the general mechanisms that rule planetary formation and evolution. We are contributing to this cause by carrying out a large programme using an array of medium-class telescopes to perform photometric follow-up of the transits of known exoplanets. The main aim of our programme is to collect high-quality transit light curves that we use to refine measurements of the physical parameters of the corresponding planetary systems in a homogeneous way \citep{mancinisouth:2016}.

During a transit event, the planet acts as an opaque mask which `scans' a stripe of the parent star's photosphere. In the case of hot Jupiters, which have a relatively large size, transiting main-sequence stars similar to the Sun, this scanning can reveal star-spots. These regions are recorded as small `bumps' in the transit light curve and provide additional information about the stellar activity and planetary orbit. High-precision, low-scatter and unbinned light curves are needed to catch these bumps, whose amplitude is colour-dependent. Thanks to observations of planetary transits, star-spots have been now detected and characterised in many circumstances \citep{rabus:2009,silva:2010,sing:2011,sanchis:2011,mohler:2013,mancini:2013c,huitson:2013,sanchis:2013,mancini:2014,mancini:2015,beky:2014}.

In this work, we study the transiting planetary system WASP-52 \citep{hebrard:2013}. This is composed of a low-density, inflated hot Jupiter, WASP-52\,b (mass $M_{\rm p}\approx 0.5\,M_{\rm Jup}$ and radius $R_{\rm p}\approx 1.3\,R_{\rm Jup}$), which orbits a K2\,V star, WASP-52\,A, every 1.75\,d. \citet{hebrard:2013} observed emission cores in the Ca\,{\sc ii} H+K lines of the WASP-52 spectra, which indicate that the star is active. They also estimated its rotational period, $P_{\rm rot}=11.8\pm3.3$\,d, and a gyrochronological age of $0.4_{-0.2}^{+0.3}$\,Gyr, which suggests that the star is quite young, even though no lithium was detected in its spectra. They also spectroscopically observed a transit, detected the Rossiter-McLaughlin effect and measured the sky-projected orbital obliquity to be $\lambda=24_{~\,-9}^{\circ +17}$.

Since the star is active, it may be possible to detect star-spots during transits, and we have done so. The paper is structured as follows. The photometric follow-up observations and data reduction are described in Sect.\,\ref{sec_2}. The analysis of the light curves is presented in Sect.\,\ref{sec_3}. In Sect.\,\ref{sec_4}, we revise the main physical properties of the planetary system. In Sect.\,\ref{sec_5}, we investigate the variation of the planetary radius as function of wavelength and, finally, we summarise our results in Sect.\,\ref{sec_6}.

\section{Observations and data reduction}
\label{sec_2}

\begin{figure}
\centering
\includegraphics[width=\columnwidth]{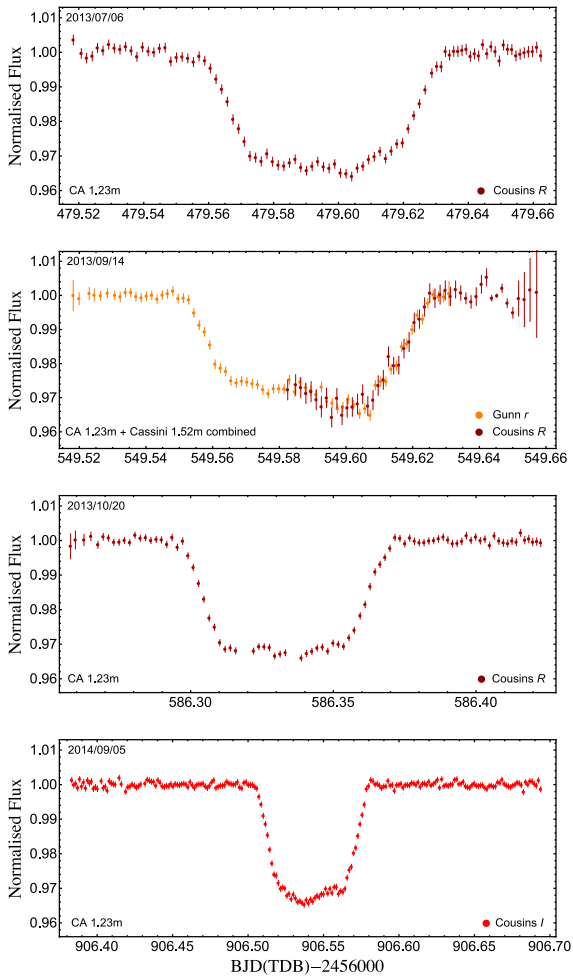}
\caption{Light curves of four transits of WASP-52\,b observed with the CA 1.23\,m telescope, shown
in date order. The second transit was partially monitored by this telescope, but fully observed with
the Cassini 1.52\,m telescope. Star-spot anomalies are visible in the second and fourth panel.}
\label{fig:lc_01}
\end{figure}

\begin{figure}
\centering
\includegraphics[width=\columnwidth]{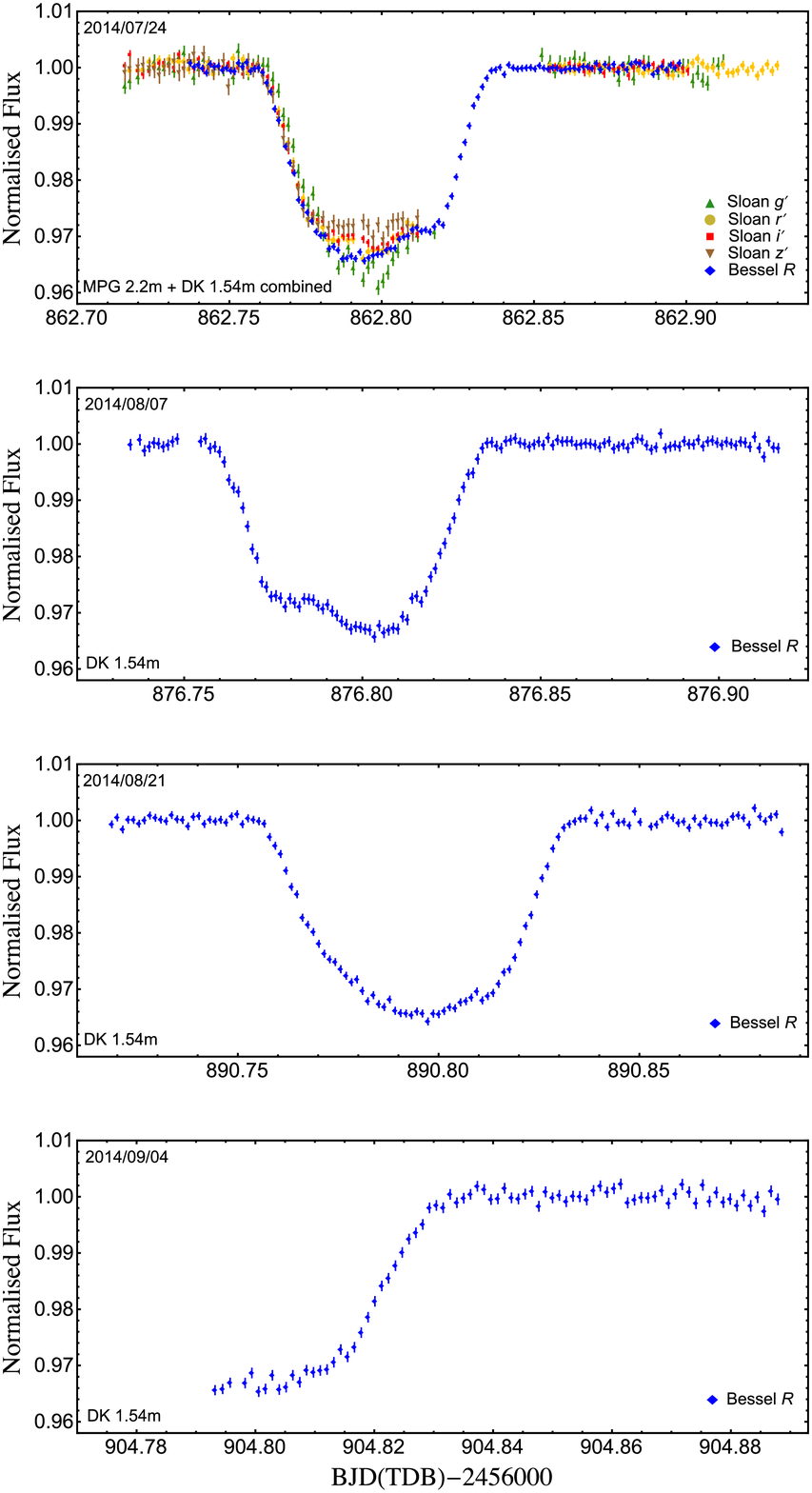}
\caption{Light curves of four (three complete and one partial) transits of WASP-52\,b observed with the
Danish 1.54\,m telescope, shown in date order. The first transit was also partially monitored by the MPG
2.2\,m telescope in four optical bands. Star-spot anomalies are visible in the first, second and third panel.}
\label{fig:lc_02}
\end{figure}

The planetary system WASP-52 is near the celestial equator so can be observed from both hemispheres. In 2013 and 2014 we monitored eight (seven complete and one partial) transits of WASP-52\,b in several optical passbands (covering $400-1000$\,nm), using four different telescopes (Table\,\ref{tab:obs}) and obtaining a total of 13 light curves. Five of the transits show anomalies that are compatible with the presence of occulted star-spots on the photosphere of the parent star (see Figs\,\ref{fig:lc_01} and \ref{fig:lc_02}).
We can actually exclude that the anomalies on the light curves are caused by plages because plages do not resize in the photosphere, but in the chromosphere. 
Actually, using a telescope working in the optical, we can only see stellar light coming from the photosphere, except for a small amount of light in the H and K lines at 3933.7 and 3968.5\,\AA, H$\alpha$ at 6562.8\,\AA, and a few other lines. That means that if WASP-52\,b occulted a plage, we would have to observe anomalies only in the $g^{\prime}$ band, i.e. the Gamma-Ray Burst Optical and Near-Infrared Detector (GROND) bluest band (see Sect.\,\ref{sec_2.4}), as in the case of HATS-2 \citep{mohler:2013}.
We can also exclude that the anomalies on the light curves are caused by bright spots (i.e. faculae) because of the transit depth: the data points that were not affected by star-spots are at the right transit depth; on the contrary, if we consider that the planet occulted hot spots, then the data points that were not affected by them are at the wrong transit depth. Moreover, faculae are mostly seen at the solar limb, not in the disc centre. So, also their effect to the transit light curves should be negligible, except maybe close to the limb, but this is not the case.

%

The transit on 2013/09/14 was simultaneously followed by two telescopes at different observing sites (Fig.\,\ref{fig:lc_01}, second panel). The complete transit was observed from Italy, and part was also observed from Spain on a cloudy night. Interestingly, the same anomalous feature is seen in both light curves, demonstrating the power of the two-site observational strategy in differentiating true astrophysical signal from systematic noise \citep{ciceri:2013,mancini:2013a}.

The transit on 2014/07/24 was also simultaneously observed using two telescopes, located on the same site in the Southern hemisphere (Fig.\,\ref{fig:lc_02}, first panel). Also in this case, one telescope monitored the entire transit but the other telescope missed half of it, due to a failure of the telescope control system. Also in this case, an anomaly was recorded by both telescopes.

Observations were all performed by \emph{defocussing} the telescopes, in order to increase the photometric precision \citep{southworth:2009}, and using autoguiding. In all cases except the MPG 2.2\,m telescope, the CCDs were windowed to decrease the readout time and therefore increase the cadence of the observations. The reduced data will be made available at the {\sc CDS}\footnote{{\tt http://cdsweb.u-strasbg.fr/}}

\begin{table*}
{\tiny
\centering
\setlength{\tabcolsep}{4pt}
\caption{Details of the transit observations presented in this work. $N_{\rm obs}$ is the number of observations,
$T_{\rm exp}$ is the exposure time, $T_{\rm obs}$ is the observational cadence, and `Moon illum.' is the geocentric
fractional illumination of the Moon at mid-night ({\sc ut}). The aperture sizes are the radii of the software apertures
for the star, inner sky and outer sky, respectively. Scatter is the \emph{rms} scatter of the data versus a fitted
model. The last column specifies if the transit was observed by two telescopes and if it was affected by star-spot anomalies.}
\label{tab:obs}
\begin{tabular}{lcccrcclcrccc} \hline
Telescope & Date of   & Start time & End time  &$N_{\rm obs}$ & $T_{\rm exp}$ & $T_{\rm obs}$ & ~~Filter & Airmass & Moon & Aperture & Scatter & Simultaneous /  \\
               & first obs &    ({\sc ut})    &   ({\sc ut})    &              & (s)           & (s)           &        &         &illum.& radii (px) & (mmag) & star-spot \\
\hline
CA\,1.23\,m & 2013/07/06 & 00:26 & 03:53 &  90~ &  90--130 & 110--150 & Cousins $R$ & $2.07 \rightarrow 1.15$  &  5\%  & 28,38,50  & 1.26 & no / no\\
Cassini\,1.52\,m & 2013/09/14 & 00:26 & 03:08 &  70~ &  120 & 137 & Gunn $r$ & $1.31 \rightarrow 2.35$  &  66\%  & 15,20,50  & 1.27 & yes / yes  \\
CA\,1.23\,m & 2013/09/14 & 01:58 & 03:46 &  48~ &  120--180 & 133--193 & Cousins $R$ & $1.29 \rightarrow 1.90$  &  66\%  & 22,32,60  & 1.79 & yes / yes  \\
CA\,1.23\,m & 2013/10/20 & 18:11 & 22:08 &  89~ &  120--150 & 120--170 & Cousins $R$ & $1.72 \rightarrow 1.14 \rightarrow 1.15$  &  96\%  & 25,35,55  & 0.87 & no / no \\
MPG\,2.2\,m & 2014/07/24 & 05:10 & 09:55 &  98~ & 110 & 140 & Sloan $g^{\prime}$ & $1.74 \rightarrow 1.27 \rightarrow 1.49$  &  ~~7\%  & 35,70,80  & 1.67 & yes / yes \\
MPG\,2.2\,m & 2014/07/24 & 05:10 & 09:55 &  94~ & 110 & 140 & Sloan $r^{\prime}$ & $1.74 \rightarrow 1.27 \rightarrow 1.49$  &  ~~7\%  & 40,60,80  & 0.57 & yes / yes \\
MPG\,2.2\,m & 2014/07/24 & 05:10 & 09:55 &  83~ & 110 & 140 & Sloan $i^{\prime}$ & $1.74 \rightarrow 1.27 \rightarrow 1.49$  &  ~~7\%  & 35,70,80  & 0.63 & yes / yes \\
MPG\,2.2\,m & 2014/07/24 & 05:10 & 09:55 &  96~ & 110 & 140 & Sloan $z^{\prime}$ & $1.74 \rightarrow 1.27 \rightarrow 1.49$  &  ~~7\%  & 50,70,90  & 1.46 & yes / yes \\
Danish\,1.54\,m & 2014/07/24 & 05:34 & 09:25 &  114~ & 100--110 & 113--123 & Bessel $R$ & $1.55 \rightarrow 1.27 \rightarrow 1.41 $  &  ~~7\%  & 20,30,50  & 0.51 & yes / yes \\
Danish\,1.54\,m & 2014/08/07 & 05:09 & 09:51 &  134~ & 100 & 113 & Bessel $R$ & $1.42 \rightarrow 1.27 \rightarrow 1.86$  &  83\%  & 19,28,50  & 0.70 & no / yes \\
Danish\,1.54\,m & 2014/08/21 & 05:27 & 09:05 &  123~ & 100 & 113 & Bessel $R$ & $1.31 \rightarrow 1.27 \rightarrow 1.95$  &  16\%  & 19,28,50  & 0.70 & no / yes \\
Danish\,1.54\,m & 2014/09/04 & 07:02 & 09:18 &   82~ & 90--100 & 103--113 & Bessel $R$ & $1.33 \rightarrow 2.24$  &  70\%  & 20,30,50  & 0.74 & no / no \\
CA\,1.23\,m & 2014/09/05 & 21:11 & 04:39 & 217~ &  100--110 & 113--123 & Cousins $I$ & $1.73 \rightarrow 1.14 \rightarrow 2.15$  &  87\%  & 19,28,50  & 0.74 & no / yes \\
\hline
\end{tabular}
}\end{table*}

\subsection{CA 1.23\,m telescope}
\label{sec_2.1}

Four transits of WASP-52\,b were monitored with the Zeiss 1.23\,m telescope at the German-Spanish Astronomical Center at Calar Alto in Spain. This telescope has a focal length of 9857.1\,mm and is equipped with the DLR-MKIII camera, which has 4k $\times$ 4k pixels of size 15\,$\mu$m. The plate scale is 0.32 arcsec\,pixel$^{-1}$ and the field-of-view (FOV) is 21.5\,arcmin $\times$ 21.5\,arcmin. The first three transits were observed in 2013 through a Cousins-$R$ filter, whereas the last one was observed in 2014 through a Cousins-$I$ filter and clearly exhibits an anomaly caused by stellar activity. Three transits were completely observed, and one was partially observed due to unfavourable weather conditions. The resulting light curves are plotted in Fig.\,\ref{fig:lc_01}.

\subsection{Cassini 1.52\,m telescope}
\label{sec_2.2}

The partial transit event recorded with the CA 1.23\,m telescope on 2014 September 14, was completely observed with the Cassini 1.52\,m telescope from the Astronomical Observatory of Bologna in Loiano, Italy. This telescope has a focal length of 12\,m, a focal ratio of $f/8$ and is equipped with the BFOSC (Bologna Faint Object Spectrograph and Camera) imager, which has a back-illuminated CCD with $1300 \times 1340$\,pixels and a pixel size of $20\,\mu$m. With the focal reducer the telescope is a $f/5$, so that the current plate scale is 0.58 arcsec pixel$^{-1}$ and the FOV is 13\,arcmin $\times$ 12.6\,arcmin. A Gunn-$r$ filter was used, and the light curve is plotted in the second panel of Fig.\,\ref{fig:lc_01}. It shows an anomaly compatible with a star-spot complex occulted by the planet during the transit event. The shape of the second part of the light curve is very similar to that observed from Calar Alto.

\subsection{Danish 1.54\,m telescope}
\label{sec_2.3}

Four transits (three complete and one partial) of WASP-52\,b were observed through a Bessel-$R$ filter between 2014 July and September with the DFOSC (Danish Faint Object Spectrograph and Camera) imager mounted on the Danish 1.54\,m Telescope at ESO La Silla, Chile. Since 2012 DFOSC has been equipped with a new camera, an e2v CCD with $2048\times4096$ pixels and 32-bit encoding. The plate scale is 0.39 arcsec\,pixel$^{-1}$. In the current optical set-up, the incoming light illuminates only half of the CCD, so the FOV is 13.7 arcmin $\times$ 13.7 arcmin. Once again, the three complete light curves present anomalies compatible with parent star's star-spot activity (see Fig.\,\ref{fig:lc_02}). Unfortunately, the transit recorded on 2014/09/04 was not fully covered due to technical problems at the beginning of the observations.

\subsection{MPG 2.2\,m telescope}
\label{sec_2.4}

The transit event recorded with the Danish 1.54\,m telescope on 2014/07/24 was partially monitored with the MPG 2.2\,m telescope, located at the same observatory. This telescope has a focal length of 17.6\,m and mounts three different instruments. We used GROND, an imaging camera with the ability to observe in four optical (similar to Sloan $g^{\prime}, r^{\prime}, i^{\prime}, z^{\prime}$) and three near-IR (NIR) bands ($J$, $H$, $K$) simultaneously. Since the photometric precision of the NIR arms is not as good as that of the optical ones \citep{pierini:2012,mancini:2013b,mancini:2014a,nikolov:2013,chen:2014}, we only considered the optical data. Incoming light is splitted by dichroics in the optical arms and channelled towards four back-illuminated  $2048\times2048$ pixel E2V CCDs. The pixel size of the CCDs is 13.5\,$\mu$m, the plate scale is 0.158\,arcsec\,pixel$^{-1}$, and the FOV is 5.4\,arcmin $\times$ 5.4\,arcmin. Due to a critical failure of the telescope control system, the observations were interrupted for $\sim45$\,minutes, so the egress phases of the transit were not observed (see panel 1 of Fig.\,\ref{fig:lc_02}). Incomplete light curves are more difficult to model, especially if they are affected by anomalies. Moreover, the amplitude of a star-spot anomaly is colour-dependent. Indeed, star-spots have a lower temperature with respect to the photosphere and, therefore, the flux ratio is expected to be lower in the blue than in red. This implies that moving from $g^{\prime}$ to $z^{\prime}$ stars-pots become brighter and star-spot features on transit light curves become less evident. In our case, the transit observed with GROND was interrupted during the star-spot anomaly and this makes the modelling quite difficult. We found that the GROND $r^{\prime}$ light curve is in a good agreement with that observed with the Danish telescope in Bessel $R$, but the same is not true for the other bands. In particular the $g^{\prime}$ band looks to be affected by a systematic or a brighter spot, while the $i^{\prime}$ and $z^{\prime}$ bands are quite flat during totality.

\subsection{Aperture photometry}
\label{sec_2.5}

All the data were reduced in a homogeneous way, using a revised version of the {\sc defot} code \citep{southworth:2009,southworth:2014}. In brief, the scientific images were calibrated by means of master-bias and master-flat frames and then their two-dimensional offset with respect to a reference frame was calculated. We performed standard aperture photometry to extract the light curves of the transits. This was done by running the {\sc aper} routine, after having placed the three apertures by hand on the target and on a set of good comparison stars. The sizes of the apertures were decided after several attempts, by selecting those having the lowest scatter when compared with a fitted model. The resulting light curves were normalised to zero magnitude by fitting a straight line to the out-of-transit data. As in our previous papers, we enlarged the uncertainties for each light curve, as they are generally underestimated in the aperture-photometry process. This enlargement of the error bars was performed by imposing a reduced $\chi^2$ of $\chi_{\nu}^2=1$ versus a fitted model\footnote{This was done for each light curve individually. This approach does not fully capture correlated noise, but correlated noise is accounted for in the error bars of the final photometric parameters, because these are obtained from the parameters calculated from each light curve independently.
}.
The final light curves are plotted in Figs\ \ref{fig:lc_01} and \ref{fig:lc_02}.

\section{Light-curve analysis}
\label{sec_3}

\begin{figure*}
\centering
\includegraphics[width=18.0cm]{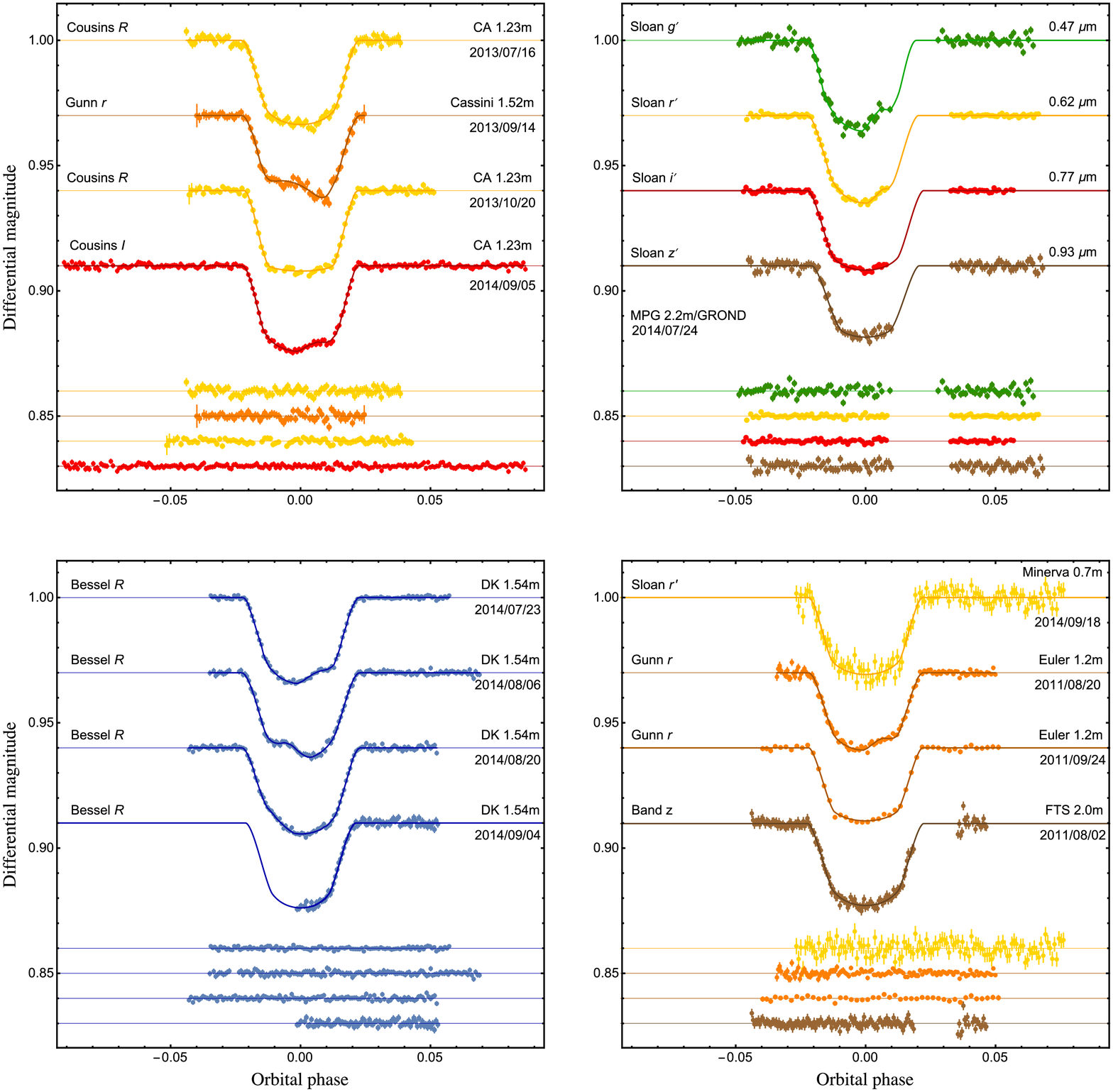}
\caption{The light curves of WASP-52 used in the analysis of the physical parameters of the system. They are
plotted versus orbital phase and are compared to the best-fitting models. The residuals of the fits are shown
at the base of each panel. The first three panels refer to the new light curves presented in this work, while
the fourth panel contains light curves taken from the literature and re-examined in our study. Labels indicate
the observation date, the telescope and the filter that were used for each data set.}
\label{fig:fit_01}
\end{figure*}

\begin{figure}
\centering
\includegraphics[width=\columnwidth]{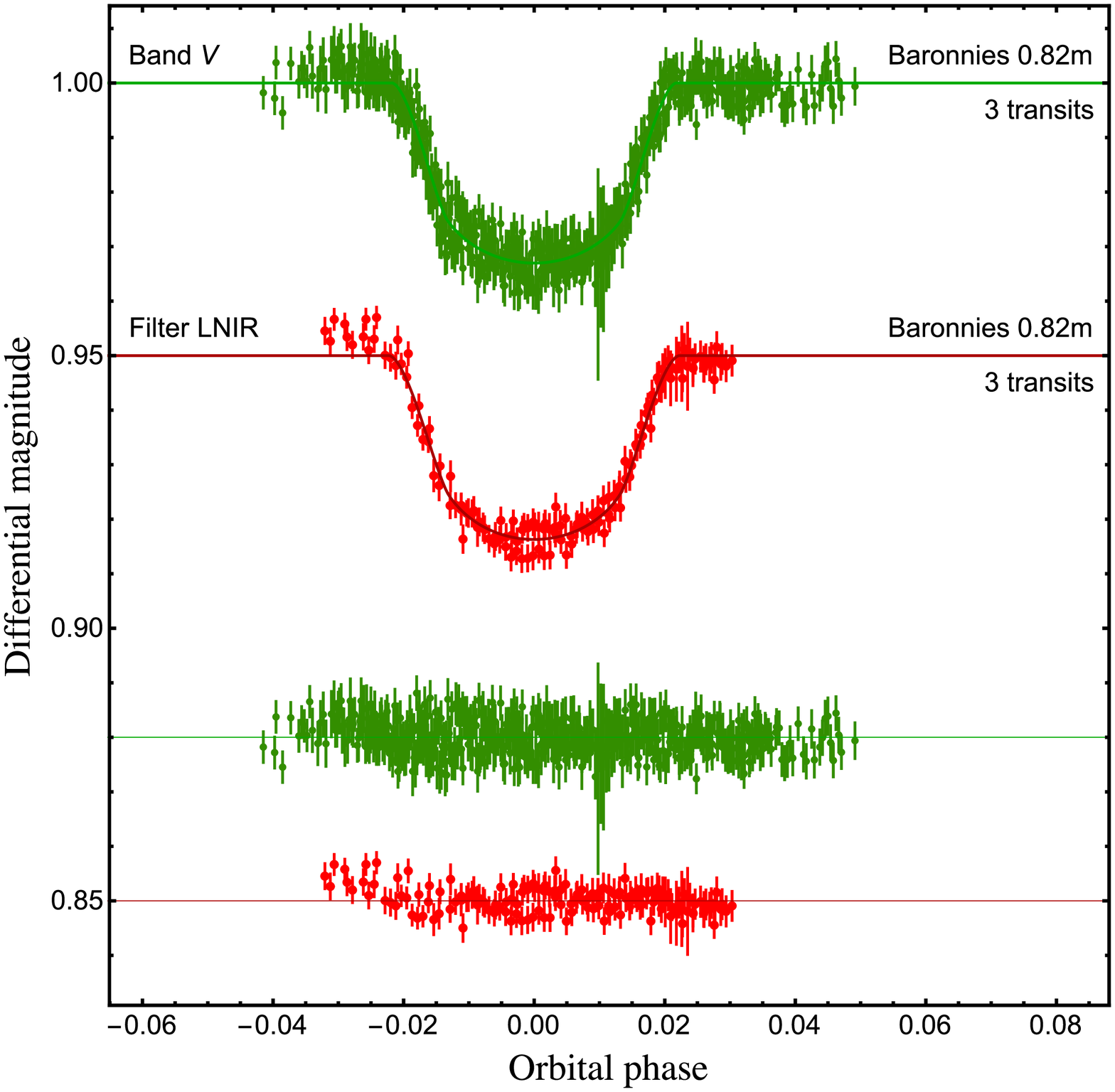}
\caption{Phase-folded binned light curves of six transits of WASP-52 observed with the Baronnies 0.82\,m telescope, three through
a $V$ filter and three through an Astrodon Luminance Near Infrared (LNIR) filter. The data are available on the ETD web archive.}
\label{fig:fit_02}
\end{figure}

Since most of our light curves of the transits of WASP-52\,b present star-spot crossing events, we have modelled them with a code designed for this task.
From secular observations of the Sun, we know that star-spots can appear as big, single circular spots, or as a complex of several spots with different sizes.
However, the quality and the sampling of the data are usually not sufficient for detecting fine structures in the star-spot anomalies. Therefore, it is normal practice to models these anomalies in the transit light curves as single circular star-spots. As in previous cases \citep{mancini:2013c,mancini:2014}, we utilised the {\sc prism}\footnote{Planetary Retrospective Integrated Star-spot Model.} and {\sc gemc}\footnote{Genetic Evolution Markov Chain.} codes \citep{tregloan:2013,tregloan:2015}, which allowed us to fit both the full transit event and the shorter star-spot-occultation event simultaneously. One of the main advantages of {\sc prism+gemc} is that the user decides how many star-spots will be fitted, based on a visual inspection of each light curve. Then, each star-spot complex is modelled as a circular spot with these parameters: the longitude and co-latitude of its centre ($\theta$ and $\phi$), its angular radius ($r_{\mathrm{spot}}$) and its contrast ($\rho_{\mathrm{spot}}$), which is the ratio of the surface brightness of the star-spot with respect to the surrounding photosphere. At the same time, the geometrical parameters that are fitted are the sum and the ratio of the fractional radii ($r_{\mathrm{A}}+r_{\mathrm{b}}$ and $k=r_{\mathrm{b}}/r_{\mathrm{A}}$, where the fractional radii are defined as $r_{\mathrm{A}} = R_{\star}/a$ and $r_{\mathrm{b}} = R_{\mathrm{p}}/a$, where $R_{\star}$ and $R_{\mathrm{p}}$ are the true radii of the star and planet, and $a$ is the orbital semimajor axis), the orbital inclination ($i$), the orbital period ($P$), the time of transit midpoint ($T_{0}$), and the coefficients of the quadratic limb darkening law ($u_{\mathrm{A}}$ and $v_{\mathrm{A}}$). We assumed a circular orbit \citep{hebrard:2013}. Each of our transit light curves with anomalies was modelled considering a single star-spot complex.

We also analysed the best three light curves presented in the discovery paper (we excluded incomplete and low-quality light curves) and taken with the Euler 1.2\,m and the FTS 2\,m telescopes \citep{hebrard:2013}, one light curve obtained with the Minerva 0.7\,m telescope \citep{swift:2015}, and six taken with the Baronnies 0.82\,m telescope (available on the ETD\footnote{The Exoplanet Transit Database (ETD) can be found at {\tt http://var2.astro.cz/ETD}.} web archive). The last were grouped according to the filter used. Details are reported in Table\,\ref{Table:fits}. The first Euler light curve presents a star-spot anomaly and we modelled it with {\sc prism+gemc}. The other light curves do not show detectable anomalies, so we reanalysed them with a much faster code, {\sc jktebop}\footnote{The {\sc jktebop} code is available at \\ {\tt http://www.astro.keele.ac.uk/jkt/codes/jktebop.html}.} (see \citealp{southworth:2013} and references therein), which fitted the same photometric parameters as {\sc prism+gemc} except for the spot parameters.

The results of all the fits are summarised in Table\,\ref{Table:fits} and displayed in Figs.\ \ref{fig:fit_01} and \ref{fig:fit_02}. The values of the photometric parameters ($r_{\rm A}+r_{\rm b}$, $k$ and $i$) were combined into weighted means to get the final values. The best-fitting parameters of the star-spots are reported in Table\,\ref{Table:fits2}. 
The limb-darkening coefficients were fitted during the fits and in most of the cases the values agree with the theoretical ones within the uncertainties.

\begin{table*}
\caption{Parameters of the {\sc prism}+{\sc gemc} and {\sc jktebop} best fits of the WASP-52 light curves used in this work.
The final parameters, given in bold, are the weighted means of the results for the individual data sets. Results from the discovery paper are included at the base of the table for comparison.}
\label{Table:fits}
\centering
\scriptsize
\begin{tabular}{llllccccc}
\hline
Telescope & date & \,\,\,\,filter & code & $r_{\mathrm{A}}+r_{\mathrm{b}}$ & $r_{\mathrm{b}}/r_{\mathrm{A}}$ & $i^{\circ}$ & $u_{\mathrm{A}}$ & $v_{\mathrm{A}}$ \\
\hline
\multicolumn{5}{l}{\textbf{Light curves from this work}} \\
CA 1.23\,m  & 2013/07/05 & Cousins   $R$ & {\sc gemc} & $0.1629 \pm 0.0050$ & $0.1660 \pm 0.0028$ & $85.26 \pm 0.48$ & $0.24 \pm 0.16$ & $0.48 \pm 0.25$ \\
CA 1.23\,m  & 2013/10/20 & Cousins   $R$ & {\sc gemc} & $0.1587 \pm 0.0032$ & $0.1680 \pm 0.0013 $ & $85.19 \pm 0.30$ & $0.13 \pm 0.08$ & $0.16 \pm 0.14$ \\
CA 1.23\,m  & 2014/09/05 & Cousins   $I$ & {\sc gemc} &  $0.1568 \pm 0.0020$ & $0.1640 \pm 0.0025 $ & $85.85 \pm 0.26$ & $0.37 \pm 0.14$ & $0.37 \pm 0.25$ \\[2pt]
Cassini 1.52\,m  & 2013/09/13 & Gunn   $r$ & {\sc gemc} & $0.1690 \pm 0.0077$ & $0.1704 \pm 0.0065 $ & $84.68 \pm 0.77$ & $0.45 \pm 0.26$ & $0.34 \pm 0.24$   \\[2pt]
Danish 1.54\,m  & 2014/07/24 & Bessel $R$ & {\sc gemc} &  $0.1605 \pm 0.0010$ & $0.1643 \pm 0.0015 $ & $85.40 \pm 0.14$ & $0.62 \pm 0.08$ & $0.19 \pm 0.13$ \\
Danish 1.54\,m  & 2014/08/06 & Bessel $R$ & {\sc gemc} &  $0.1666 \pm 0.0036$ & $0.1646 \pm 0.0023 $ & $84.97 \pm 0.33$ & $0.66 \pm 0.20$ & $0.24 \pm 0.22$ \\
Danish 1.54\,m  & 2014/08/20 & Bessel $R$ & {\sc gemc} &  $0.1622 \pm 0.0020$ & $0.1660 \pm  0.0016 $ & $85.30 \pm 0.20$ & $0.53 \pm 0.06$ & $0.12 \pm 0.10$ \\
[2pt]
MPG 2.2\,m & 2014/07/24 & Sloan $g^{\prime} $ & {\sc gemc} &  $0.1584 \pm 0.0135$ &  $0.1616 \pm 0.0073$ & $86.22 \pm 1.20$ & $0.58 \pm 0.22$ & $0.28 \pm 0.21$ \\
MPG 2.2\,m & 2014/07/24 & Sloan $r^{\prime} $ & {\sc gemc} &  $0.1622 \pm 0.0024$ &  $0.1631 \pm 0.0011$ & $85.42 \pm 0.18$ & $0.65 \pm 0.02$ & $0.33 \pm 0.02$ \\
\hline
\multicolumn{5}{l}{\textbf{Light curves from the literature}} \\
FTS 2.0\,m &   2011/08/02 & Gunn $r$ & {\sc jktebop} & $0.1644 \pm 0.0054$ & $0.1653 \pm 0.0026 $ & $84.97 \pm 0.47$ & $0.53 \pm 0.10$ & $0.13 \pm 0.06$\\[2pt]
Euler 1.2\,m &   2011/08/20 & Gunn $r$ & {\sc gemc}    & $0.1621 \pm 0.0017$ & $0.1610 \pm 0.0013 $ & $84.81 \pm 0.10$ & $0.48 \pm 0.02$ & $0.12 \pm 0.07$\\
Euler 1.2\,m &   2011/09/24 & Gunn $r$ & {\sc jktebop} & $0.1645 \pm 0.0041$ & $0.1588 \pm 0.0016 $ & $84.70 \pm 0.39$ & $0.37 \pm 0.12$ & $0.13 \pm 0.06$\\[2pt]
Minerva 0.7\,m & 2014/09/18 & Sloan $r^{\prime} $ & {\sc gemc} &  $0.1587 \pm 0.0165$ &  $0.1596 \pm 0.0092$ & $85.32 \pm 1.64$ & $0.41 \pm 0.34$ & $0.21 \pm 0.06$ \\[2pt]
Baronnies 0.82\,m  & Binned data & $\,\,\,\,\,\,\,V$ & {\sc jktebop} &  $0.1589 \pm 0.0073$ & $0.1653 \pm 0.0047 $ & $85.47 \pm 0.70$ & $0.62 \pm 0.11$ & $0.06 \pm 0.07$ \\[2pt]
Baronnies 0.82\,m  & Binned data & LNIR & {\sc jktebop} &  $0.1620 \pm 0.0066$ & $0.1625 \pm 0.0044 $ & $85.35 \pm 0.64$ & $0.63 \pm 0.12$ & $0.16 \pm 0.07$ \\
\hline
{\bf Final results} & &  &  & $\mathbf{ 0.16101 \pm 0.00065}$ & $\mathbf{0.16378 \pm 0.00050}$ & $\mathbf{85.15 \pm 0.06}$ & & \\
\hline
\citet{hebrard:2013} & & & & & $0.1646 \pm 0.0012$ & $85.35 \pm 0.20$ & &   \\
\hline
\end{tabular}
\end{table*}

\begin{table*}
\centering 
\label{Table:fits2}
\caption{Star-spot parameters derived from the {\sc prism}+{\sc
gemc} fits of the transit light curves presented in this work.
\newline{$^{a}$The longitude of the centre of the spot is defined to be $0^{\circ}$ at the centre of the stellar disc and can vary from $-90^{\circ}$ to $90^{\circ}$.
$^{b}$The co-latitude of the centre of the spot is defined to be $0^{\circ}$ at the north pole and $180^{\circ}$ at the south pole.
$^{c}$Angular radius of the star-spot (note that an angular radius of $90^{\circ}$ covers half of stellar surface).
$^{d}$Spot contrast (note that 1.0 equals the brightness of the surrounding photosphere).
$^{e}$The temperatures of the star-spots are obtained by considering the photosphere and the star-spots as black bodies.}}
\begin{tabular}{lclrcccc}
\hline
Telescope & date & ~\,filter & $\theta (^{\circ})\,^{a}$~~~ & $\phi(^{\circ})\,^{b}$ & $r_{\rm spot}(^{\circ})\,^{c}$ & $\rho_{\rm spot}\,^{d}$ & Temperature (K)$\,^{e}$   \\
\hline
Euler 1.2\,m    & 2011/08/20 & Gunn $r$           & $22.83 \pm 3.48$  & $45.66 \pm 6.65 $ & $7.27 \pm 4.89$  & $0.66 \pm 0.10$ & $4572 \pm 168$ \\
Cassini 1.52\,m & 2013/09/14 & Gunn $r$           & $-13.26 \pm 2.45$ & $37.43 \pm 8.10 $ & $24.84 \pm 3.28$ & $0.71 \pm 0.09$ & $4642 \pm 151$ \\
MPG 2.2\,m      & 2014/07/24 & Sloan $g^{\prime}$ & $22.04 \pm 6.42$  & $46.36 \pm 10.01$ & $15.58 \pm 5.93$ & $0.48 \pm 0.23$ & $4467 \pm 320$ \\
MPG 2.2\,m      & 2014/07/24 & Sloan $r^{\prime}$ & $23.46 \pm 1.43$  & $46.50 \pm 2.99 $ & $8.71 \pm 1.25$  & $0.69 \pm 0.09$ & $4632 \pm 145$ \\
Danish 1.54\,m  & 2014/07/24 & Bessel $R$         & $21.84 \pm 1.50$  & $43.57 \pm 6.07 $ & $12.34 \pm 2.92$ & $0.78 \pm 0.07$ & $4735 \pm 130$ \\
Danish 1.54\,m  & 2014/08/06 & Bessel $R$         & $-16.45 \pm 2.22$ & $42.40 \pm 8.52 $ & $14.11 \pm 3.00$ & $0.72 \pm 0.20$ & $4660 \pm 278$ \\
Danish 1.54\,m  & 2014/08/20 & Bessel $R$         & $-43.77 \pm 2.40$ & $45.57 \pm 3.21 $ & $14.77 \pm 1.78$ & $0.80 \pm 0.06$ & $4757 \pm 122$ \\
CA 1.23\,m      & 2014/09/05 & Cousins $I$        & $18.55 \pm 2.16$  & $41.52 \pm 7.48 $ & $14.18 \pm 4.50$ & $0.74 \pm 0.11$ & $4631 \pm 192$ \\
\hline
\end{tabular}
\end{table*}

\subsection{Orbital period determination} %
\label{sec_3.1}

\begin{figure*}
\centering
\includegraphics[width=17.0cm]{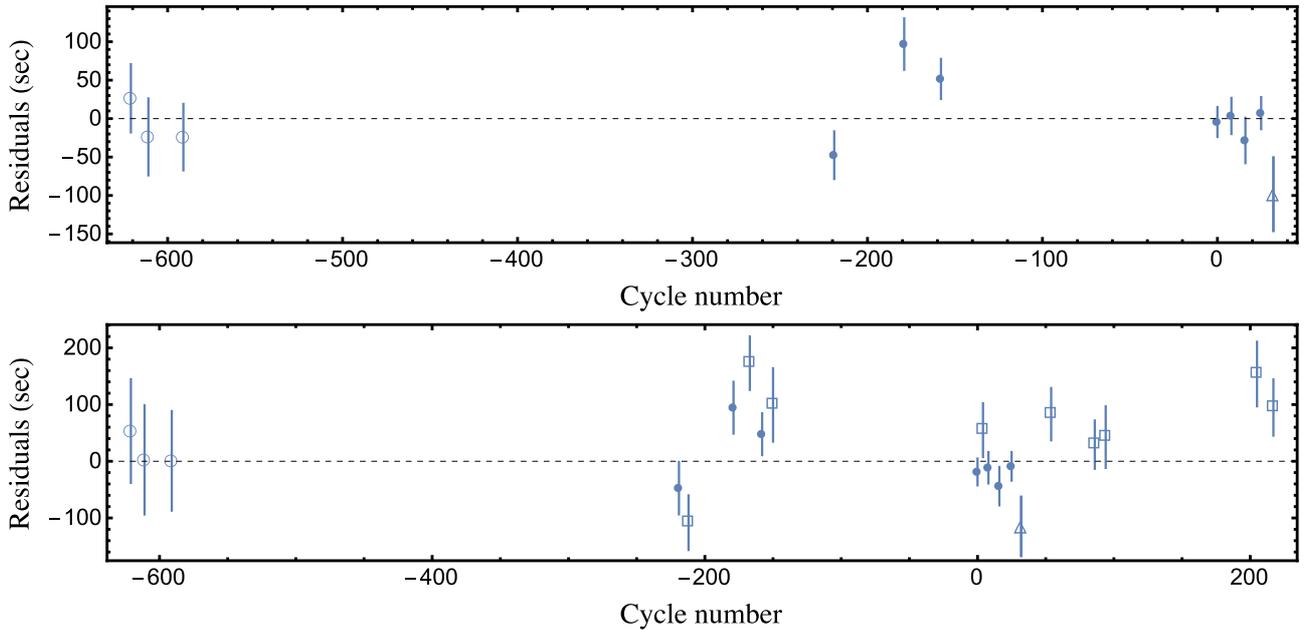}
\caption{Top panel: residuals of the times of mid-transit versus a linear ephemeris. The mid-transit times
were estimated by using the {\sc jktebop} and {\sc prism+gemc} codes. The timings from the discovery paper
\citep{hebrard:2013} are plotted using open circles, from \citet{swift:2015} with a triangle, and those based on
our observations with filled circles. We considered only single and complete light curves. Bottom panel:
similar to the upper panel, but with the addition of timings taken from the ETD archive (open boxes).}
\label{fig:OC}
\end{figure*}

\begin{table}
\caption{Times of transit midpoint of WASP-52\,b and their residuals. References:
(1) FTS 2\,m \citep{hebrard:2013};
(2) Euler 1.2\,m \citep{hebrard:2013};
(3) CA 1.23\,m (this work);
(4) Cassini 1.52\,m (this work);
(5) Danish 1.54\,m (this work);
(6) Minerva 0.7\,m \citep{swift:2015}.}
\label{tab:wasp52oc} %
\centering %
\begin{tabular}{lrrc}
\hline
Time of minimum    & Cycle & O-C~~~ & Reference  \\
BJD(TDB)$-2400000$ & no.   & (d)~~~\,     &            \\
\hline \\[-8pt]%
$55776.18395 \pm 0.00015$ & $-$621 & 0.000306 &  1   \\
$55793.68118 \pm 0.00022$ & $-$611 & $-$0.000276 &  2   \\
$55828.67680 \pm 0.00015$ & $-$591 & $-$0.000279 &  2   \\
$56479.59513 \pm 0.00016$ & $-$219 & $-$0.000551 &  3   \\
$56549.58805 \pm 0.00020$ & $-$179 & 0.001122 &  4   \\
$56586.33293 \pm 0.00012$ & $-$158 & 0.000597 &  3   \\
$56862.79771 \pm 0.00006$ & 0 & $-$0.000051 &  5   \\
$56876.79605 \pm 0.00011$ & 8 & 0.000040 &  5   \\
$56890.79393 \pm 0.00018$ & 16 & $-$0.000329 &  5   \\
$56906.54237 \pm 0.00008$ & 25 & 0.000082 &  3   \\
$56918.78962 \pm 0.00039$ & 32 & $-$0.001138 &  6   \\
\hline
\end{tabular}
\end{table}

We refined the transit ephemeris of WASP-52\,b thanks to the new photometric data. The transit times and their uncertainties were estimated using the codes mentioned above and placed on the BJD(TDB) time system. We only considered timings based on complete light curves and taken with professional telescopes. A series of very scattered points at the egress phase were excluded from the FTS 2.0\,m light curve, but this did not compromise the precision of $T_{0}$ achieved for this data set. The reference epoch was chosen as that corresponding to our best light curve, based on the {\it rms} scatter of the data (see Table\,\ref{tab:obs}). The timings were fitted with a straight line to obtain:
\begin{equation}
T_{0} = \,$BJD(TDB)$\,56862.79776(16) + 1.74978119(52)\,E , \nonumber
\end{equation}
where $E$ represents the number of orbital cycles after the reference epoch, and the quantities in brackets are the uncertainties in the two final digits of the preceding number. All transit times and their residual versus the fitted ephemeris are reported in Table\,\ref{tab:wasp52oc}. The residuals are also plotted in the top panel of Fig.\,\ref{fig:OC}. The reduced $\chi^{2}$ of the fit\footnote{The reduced $\chi^{2}$ is simply the chi-squared divided by the number of degrees of freedom. The number of degrees of freedom is given by $N-n$, where $N$ is the number of observations, and $n$ is the number of fitted parameters.} is quite high, $\chi_{\nu}^{2}= 8.98$, suggesting that the linear ephemeris does not give a good match to the observations. The uncertainties given above have been inflated to account for this by multiplying them by $\sqrt{\chi_\nu^2}$.

We then added more timings to our sample, taken from the ETD archive. We selected light curves having a complete transit coverage and a Data Quality index $\leq 1$. Adding these nine timings to the sample, we repeated the analysis, obtaining a worse $\chi_{\nu}^{2}$ of $19.60$. The new residuals are shown in the bottom panel of Fig.\,\ref{fig:OC}.

This suggests that the orbital period of WASP-52\,b is not constant, and that transit timing variations (TTVs) occur due to the presence of additional bodies in the system. However, based on our extensive experience in such kind of analysis, an excess $\chi_{\nu}^{2}$ is often caused by underestimation of the uncertainty of the measurements, which have been collected using multiple telescopes, instruments and time sources.
The presence of star-spots can also cause an excess $\chi_{\nu}^{2}$ \citep{barros:2013,oshagh:2013,ioannidis:2016}. Actually, a Lomb--Scargle periodogram of the timing residuals does not reveal a significant periodic variation. Moreover, the timing measurements are often separated by hundreds of days so are insensitive to many periodicities. The systematic observation of many subsequent transits, preferably performed with the same telescope, would be the only way to claim a TTV with a high level of confidence.

\subsection{star-spot analysis} %
\label{sec_3.2}

\begin{figure*}%
{{\includegraphics[width=5.4cm]{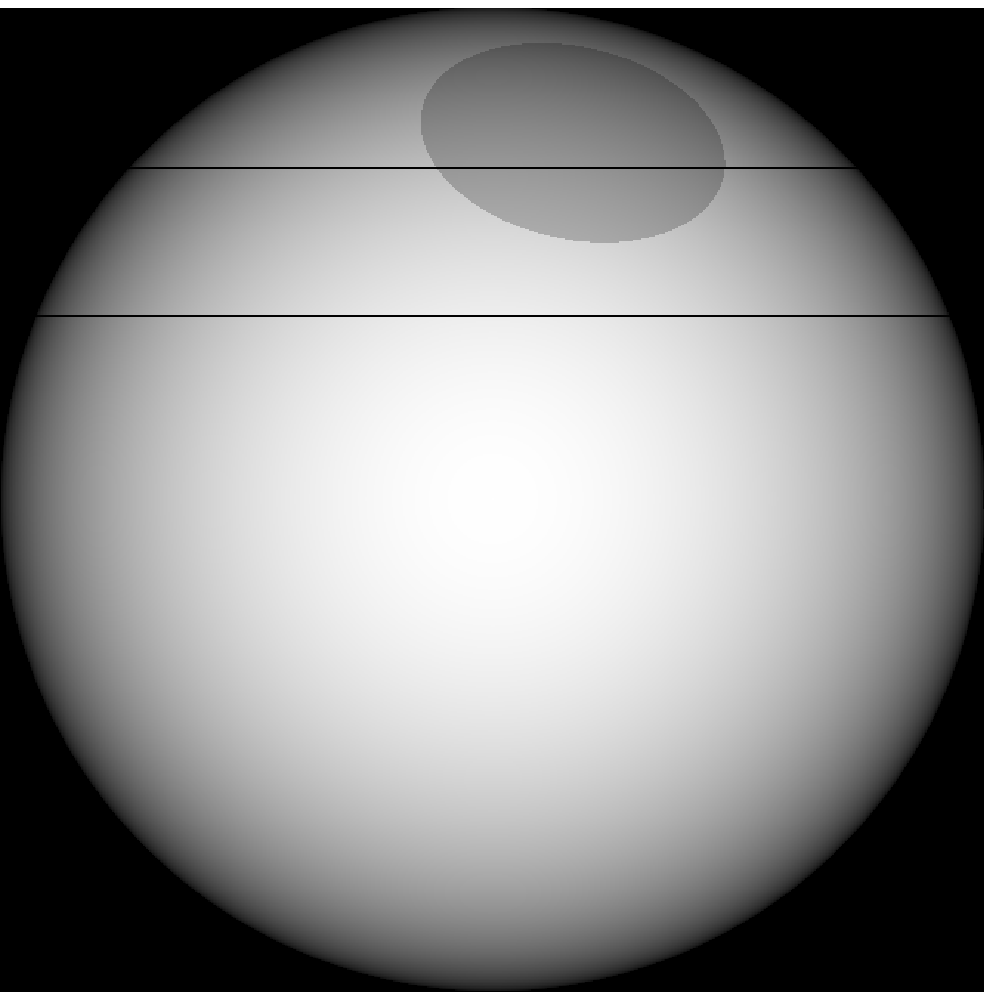} }}%
\qquad
{{\includegraphics[width=5.4cm]{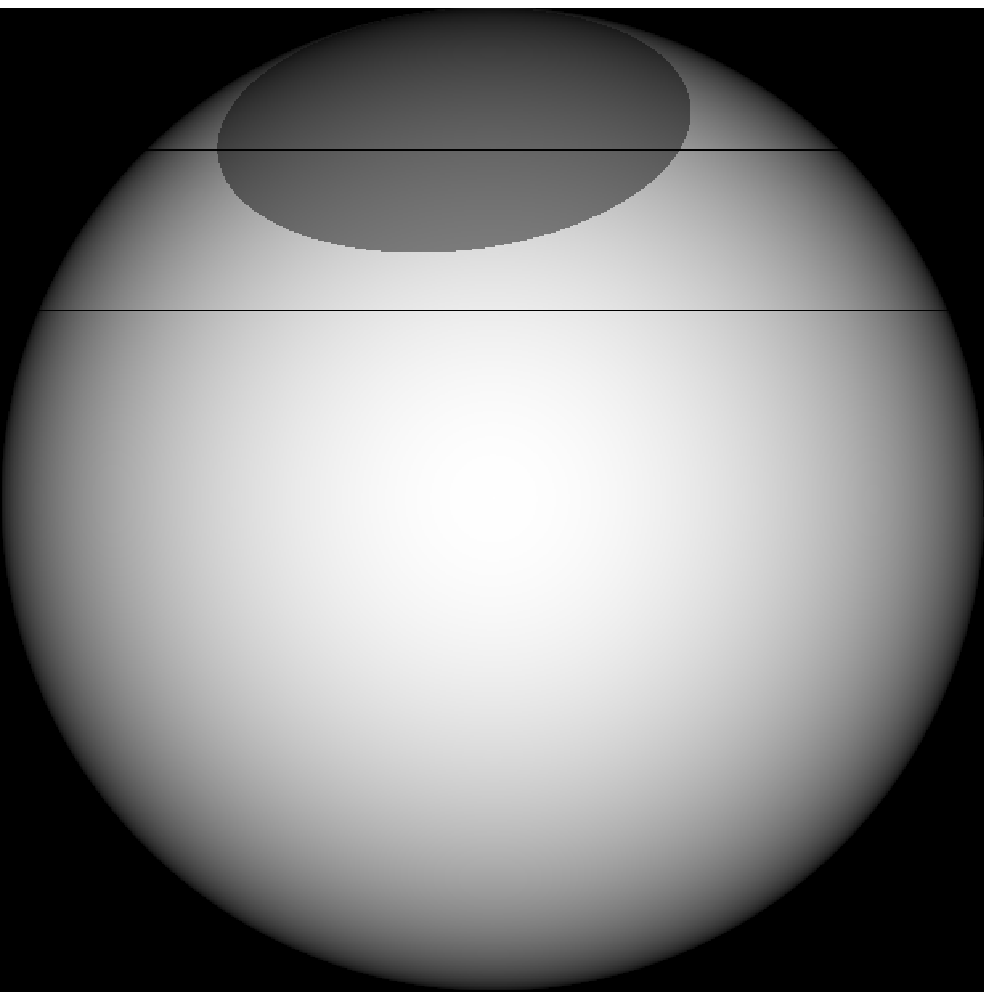} }}%
\qquad
{{\includegraphics[width=5.4cm]{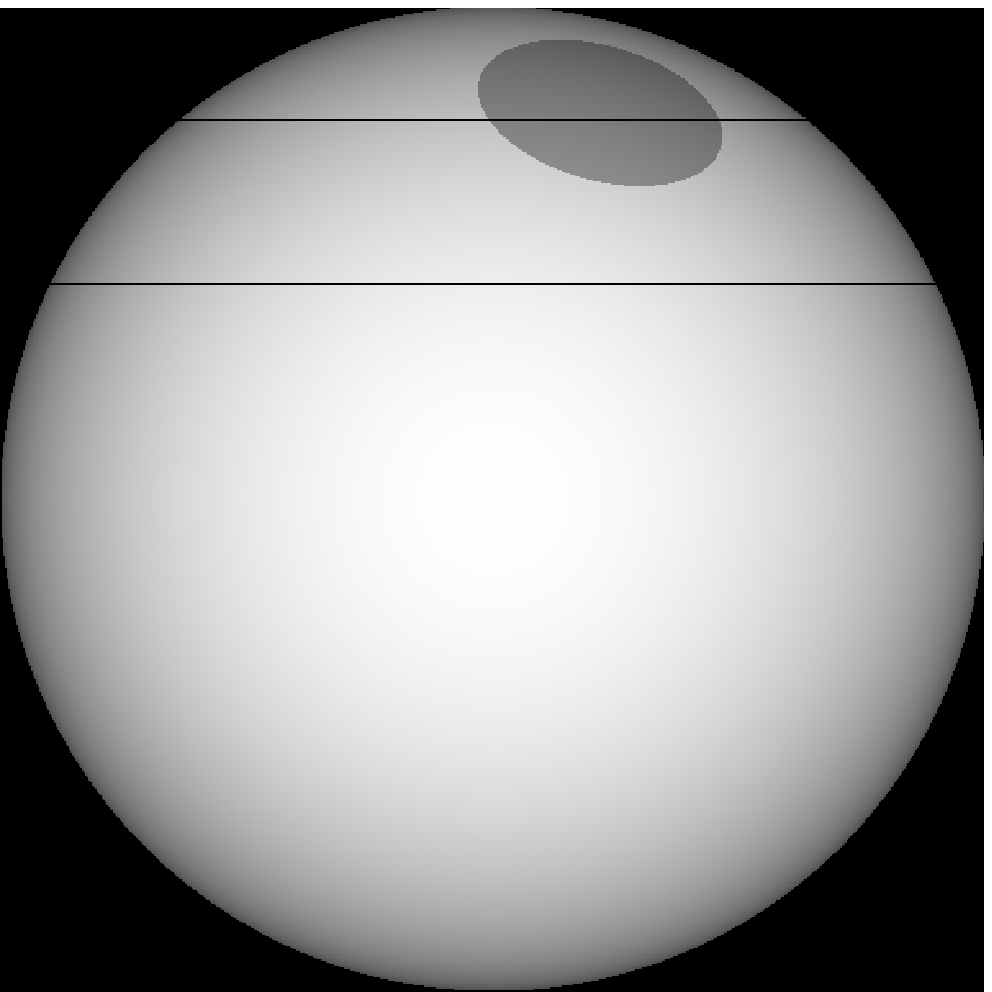} }}%
{{\includegraphics[width=5.4cm]{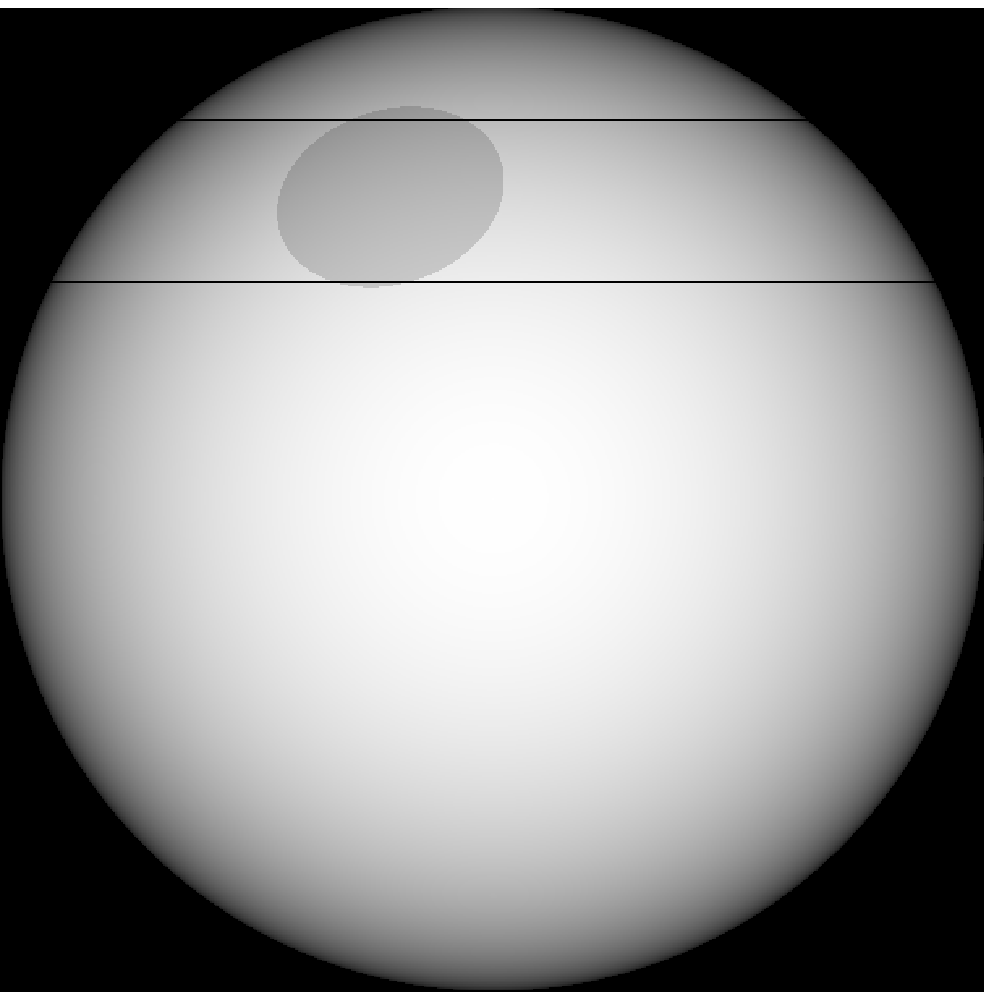}}}%
\qquad
{{\includegraphics[width=5.4cm]{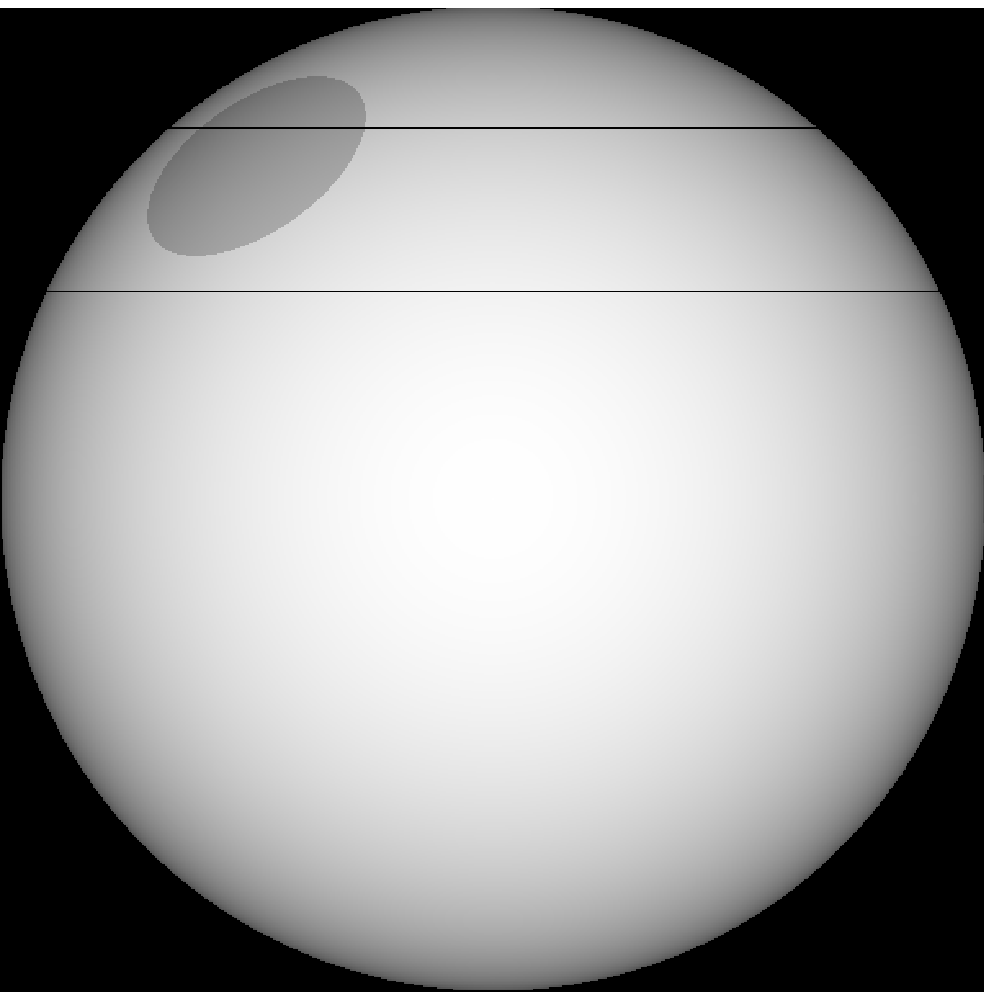}}}%
\qquad
{{\includegraphics[width=5.4cm]{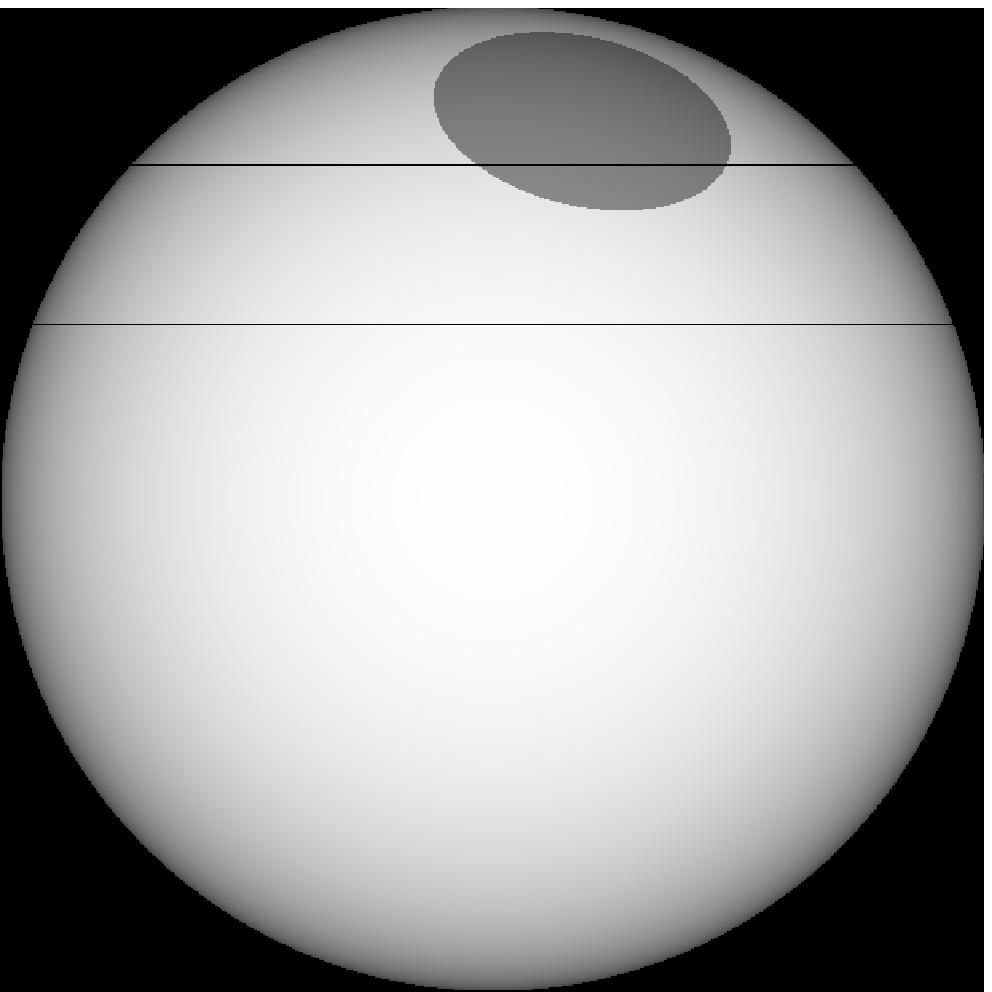}}}%
\caption{Representation of the stellar disc, star-spot position, and transit chord for the transit events
with star-spot crossings. The grey-scale of each star-spot is related to its contrast. The two horizontal lines on each panel represent the upper and lower parts of the planet pass.
Top-left panel: transit observed with the Euler 1.2\,m telescope on 2011/08/20 \citep{hebrard:2013};
top-middle panel: transit observed with the Cassini 1.52\,m telescope on 2013/09/14 (this work);
top-right panel: transit observed with the Danish 1.54\,m telescope on 2014/07/23, aka $t_{1}$ (this work);
bottom-left panel: transit observed with the Danish 1.54\,m telescope on 2014/08/06, aka $t_{2}$ (this work);
bottom-middle panel: transit observed with the Danish 1.54\,m telescope on 2014/08/20, aka $t_{3}$ (this work);
bottom-right panel: transit observed with the CA 1.23\,m telescope on 2014/09/05, aka $t_{4}$ (this work).}
\label{fig:sketches}%
\end{figure*}
\begin{figure}
\centering
\includegraphics[width=\columnwidth]{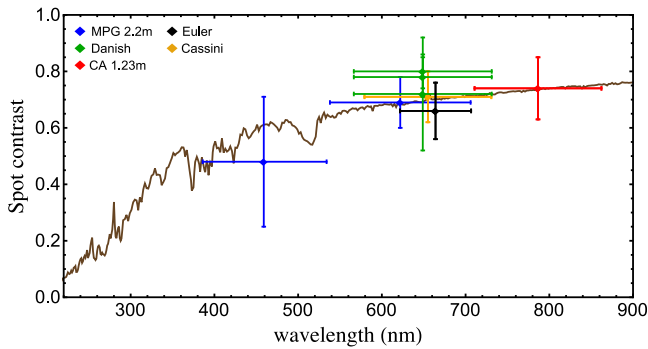}
\caption{Variation of the spot contrast with wavelength. Apart from Euler, all the points are from this
work and are explained in the plot legend. The vertical bars represent the errors in the measurements
and the horizontal bars show the FWHM transmission of the passbands used. Solid line represents the
spot contrast variation expected for a star-spot at 4650\,K over a stellar photosphere of 5000\,K, both modelled using ATLAS9 model atmospheres \citep{kurucz:1979}.}
\label{fig:spotcontrast}
\end{figure}
\begin{figure*}
\centering
\includegraphics[width=16.0cm]{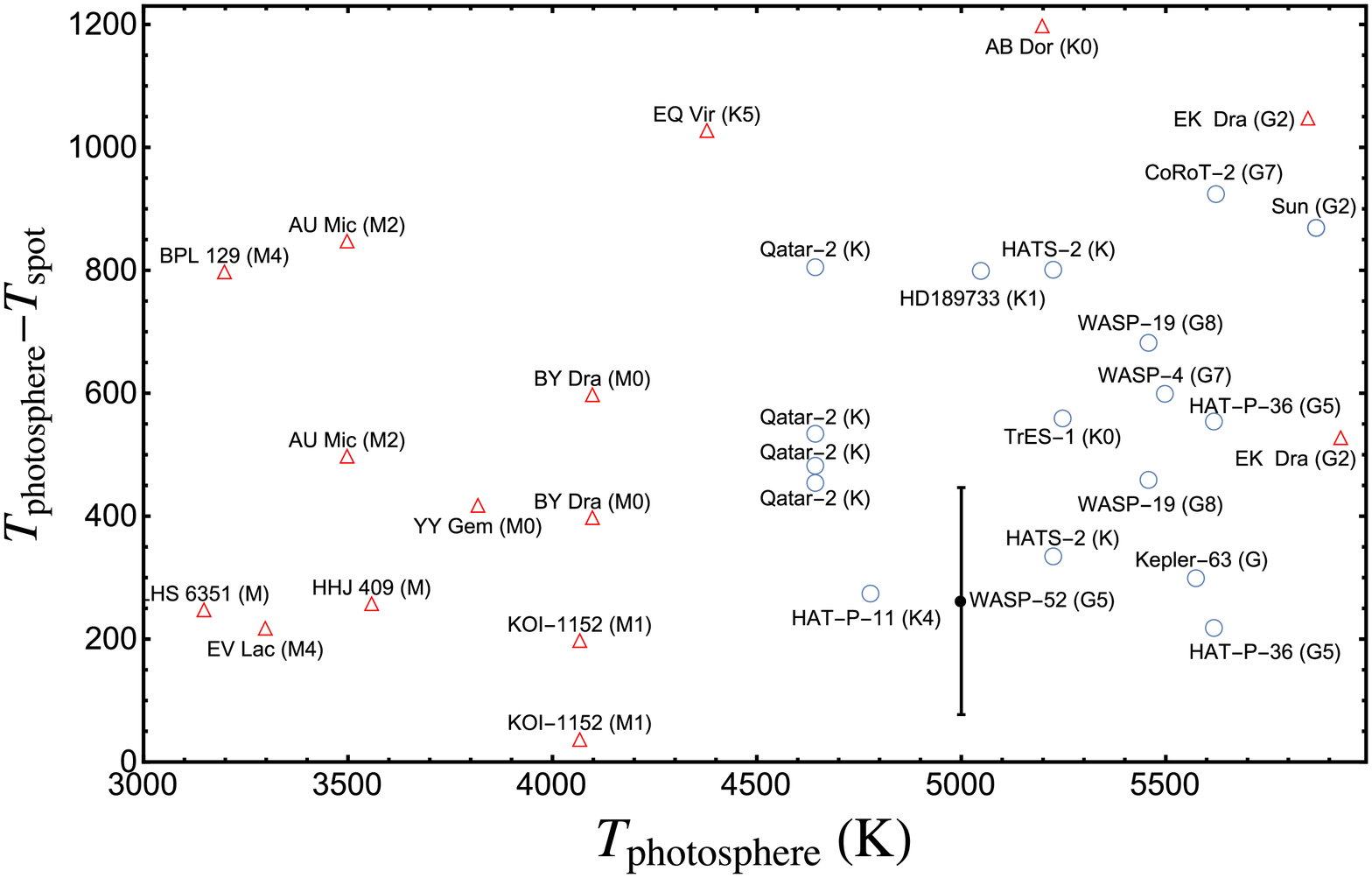}
\caption{Star-spot temperature contrast with respect to the photospheric temperature in several dwarf stars.
The name and spectral type of the star are also reported. Blue circles indicate star-spots detected during planetary transits, while red triangles were taken from \citet{andersen:2015} and refer to star-spots identified by other techniques. The references for the values are: TrES-1:
\citet{rabus:2009}, CoRoT-2: \citet{silva:2010}, HD\,189733: \citet{sing:2011}, WASP-4: \citet{sanchis:2011},
HATS-2: \citet{mohler:2013}, Kepler-63: \citet{sanchis:2013}, Qatar-2: \citet{mancini:2014}, HAT-P-36:
\citet{mancini:2015} and HAT-P-11: \citet{beky:2014}. The two values for WASP-19 are from \citet{mancini:2013c}
and \citet{huitson:2013}. Penumbra sunspot temperature was taken from \citet{berdyugina:2005}. The errorbars have been
suppressed for clarity, except for WASP-52 (black dot; this work). Note that some stars appear twice or more.}
\label{fig:star-spot_contrast}
\end{figure*}

As described above, the anomalies in some of our light curves were modelled as star-spots using the {\sc prism+gemc} codes (see Tables \ref{Table:fits} and \ref{Table:fits2}). In particular, the star-spot parameters obtained from the fit of the light curves observed simultaneously with two different telescopes on 2014/07/24 are physically consistent to each other within the uncertainties. Fig.\,\ref{fig:sketches} shows representations of the projected stellar surface with the spots and the transit chord. We note that four star-spots, of similar size, were detected in four transit events observed over 43\,d in 2014 July-September.

\subsubsection{Star-spot temperature} %
\label{sec_3.2.1}

Since star-spots have lower temperatures in comparison to stellar photosphere, we used the blackbody approximation and applied equation 1 of \citet{silva:2003}
\begin{equation}
\rho_{\rm spot}=\frac
{\exp{(h \nu / K_{\rm B} T_{\rm eff})-1}}
{\exp{(h \nu / K_{\rm B} T_{\rm spot})-1}}
\end{equation}
to estimate the temperatures of the star-spots based on their contrast (Table\,\ref{Table:fits2}), the frequency $\nu$, and on the effective temperature of WASP-52\,A, $T_{\rm eff} = 5000 \pm 100$\,K \citep{hebrard:2013}. $h$ is the Planck constant and $K_{\rm B}$ is the Boltzmann constant. The temperatures are reported in Table\,\ref{Table:fits2} and agree well with each other. This can be also seen in Fig.\,\ref{fig:spotcontrast}, in which we compare the star-spot contrasts calculated by {\sc prism+gemc} with those expected for a star-spot at 4650\,K over a stellar photosphere of 5000\,K, both modelled with ATLAS9 model atmospheres \citep{kurucz:1979}.

We then considered the three star-spots detected between 2014 July and August with the Danish telescope through the same filter (Bessel $R$), and estimated a weighted mean star-spot temperature, $4738 \pm 85$\,K. Fig.\,\ref{fig:star-spot_contrast} compares this value with those measured for other main-sequence dwarf stars.
This plot tells us that, in the case of dwarf stars, the temperature difference between the photosphere and star-spots does not seem to be correlated with spectral class. This is contrary to the trend in star-spot temperature contrasts with spectral type found by various authors, see e.g. \citet{berdyugina:2005}.

\subsubsection{Star-spot size and lifetime} %
\label{sec_3.2.2}

Over a 43\,d time interval, we have observed star-spot features in four transits of WASP-52\,b, three with the Danish 1.54\,m telescope and one with the CA 1.23\,m telescope. We label these transits as $t_1,\,t_2,\,t_3,\,t_4$ (see Table\,\ref{Table:fits} and Fig.\,\ref{fig:fit_01}). The time intervals between $t_1$ and $t_2$, and $t_2$ and $t_3$, are 14\,d, whereas that between $t_3$ and $t_4$ is 15\,d. It is worth considering if these events are due to the same star-spot, or star-spot complex, being occulted during each transit. Indeed, due to differential rotation, a large star-spot, which covers a broad latitudinal range, could possibly break into smaller spots with shorter lifetimes. First, we note that the four star-spots are located at roughly the same co-latitude and have the same angular radius within the uncertainties (Table\,\ref{Table:fits2}). Considering $t_1$, the angular size of the star-spot corresponds to a radius of $117\,737 \pm 27\,978$\,km. Knowing the size, we can estimate the lifetime of the star-spot by applying the G-W empirical relation \citep{gnevyshev:1938,waldmeier:1955}:
%
\begin{equation}
A_{\rm max}=D_{\rm GW}T,
\end{equation}
where $A_{\rm max}$ is the maximum size of the star-spot in units of micro solar hemispheres (MSH) and $T$ is the star-spot lifetime. The quantity $D_{\rm GW}$ was estimated to be $10.89 \pm 0.18$\,MSH\,d$^{-1}$ for individual sunspots \citep{petrovay:1997}. Using the G-W relationship, we estimated a lifetime of 11.7 months for the star-spot corresponding to $t_1$. However, this relation was doubted by \citet{bradshaw:2014}, who argued that it returns star-spot lifetimes that are overestimated by at least two orders of magnitude, and suggested an alternative relation, based on turbulent magnetic diffusivity operating at the supergranule scale. This new relation lowers the lifetime of the star-spot in transit $t_1$ to 70\,d for a supergranule size of $\sim 70\,000$\,km, which corresponds to 0.1\,$R_{\sun}$. Even this shorter lifetime means that the star-spot detected in $t_1$ should have lasted sufficiently long to reappear in the three subsequent transit events that we recorded.

\subsubsection{Spin-orbit alignment} %
\label{sec_3.2.3}

In general, the occultation of the same star-spot complex in two or more transit events is a clear indicator that there is a good alignment between the planet's orbital axis and its host star's spin. This allows the measurement of the sky-projected spin-orbit angle, $\lambda$, with higher precision than can normally be obtained from the Rossiter-McLaughlin effect (e.g. \citealp{tregloan:2013}). On the contrary, if we have observed two different star-spots, their latitude difference is completely degenerate with $\lambda$. Discriminating the two cases is not trivial. Crucial parameters to take into account are the rotational period of the parent star, $P_{\rm rot}$, and the difference in position and time between the star-spots.

If we monitor many transits of the same planet and have observed the same star-spot in two different transit events, then the distance $D$ covered by the star-spot, with respect to a terrestrial observer, in the time between the two detections, is given by \citep{mancini:2014}:
\begin{equation}
D=\left( n \times 2 \pi R_{\rm lat} \right)+d,
\end{equation}
where $n$ is the number of revolutions completed by the star, $R_{\rm lat}$ is the scaled stellar radius for the latitude at which the star-spot has been observed and $d$ is the arc length on the stellar photosphere between the two positions of the star-spot. Using this equation, we can calculate the rotational velocity of the star at the star-spot latitude and compare it with that measured with other techniques (for example by modelling a periodic photometric modulation in the light curve induced by a star-spot activity). This will tell us if the same star-spot may have been observed after consecutive transits or after some orbital cycles, presuming that in the latter case the star has performed one or more complete revolutions. A proper modelling and accurate analysis of the size, contrast and position of two star-spots, detected in two very close transit events, can therefore reveal that we are actually dealing with the same star-spot (e.g.\ see the case of WASP-19 discussed by \citealt{tregloan:2013}).

In our case, we are dealing with WASP-52\,A, whose rotational velocity was estimated to be $P_{\rm rot}=11.8\pm3.3$\,d \citep{hebrard:2013}. Based on the measurement of $\lambda$ by \citet{hebrard:2013}, we can exclude that WASP-52\,b is moving on a retrograde orbit. We have detected four star-spots in four different transits, in a timespan of 43\,d, which is consistent with the lifetime of a large star-spot (see Sect.\,\ref{sec_3.2.2}). For the case $n = 0$, the star would rotate unrealistically slowly. Instead, the case $n =1$ is very reasonable since it implies a rotational period of the star of $P_{\rm rot}=15.53 \pm 1.96$\,d at co-latitude of $43^{\circ}.2 \pm 4^{\circ}.9$ if we consider the star-spots detected in $t_{1}$ and $t_{2}$. This value is consistent with those coming from considering the star-spots detected in $t_{2}$ and $t_{3}$, and in $t_{3}$ and $t_{4}$, i.e.\ $P_{\rm rot}=15.10 \pm 1.27$\,d at co-latitude of $45^{\circ}.2 \pm 3^{\circ}.0$ and $13.44 \pm 1.01$\,d at a co-latitude of $44^{\circ}.9 \pm 2^{\circ}.9$, respectively. Under the assumption that we have detected the same star-spot, we can simply estimate the sky-projected angle between the stellar rotation and the planetary orbit to be $\lambda=1^{\circ}.8 \pm 21^{\circ}.8$, $6^{\circ}.6 \pm 24^{\circ}.8$, $3^{\circ}.7 \pm 9^{\circ}.9$ for the cases $t_{1}-t_{2}$, $t_{2}-t_{3}$, $t_{3}-t_{4}$, respectively. By taking the weighed mean of these values, we obtain $\lambda=3^{\circ}.8 \pm 8^{\circ}.4$, which is consistent with zero. There is a roughly $1.6\,\sigma$ disagreement between our measurement of $\lambda$ and that from (\citealt{hebrard:2013}, $\lambda=24_{~\,-9}^{\circ +17}$). Even though our result supports a very low spin-orbit misalignment, the two measurements are quite compatible.

The detection of the same star-spot in three or more transits can in principle allow the true, rather than sky-projected, orbital alignment to be found (e.g. \citealt{nutzman:2011,sanchiswinn:2011,sanchis:2013}). In the current case we have four position measurements for one star-spot, although two are at very similar longitudes so are effectively one measurement. We fitted the four positions of the spot using a coordinate transformation between the intrinsic stellar surface and the projected surface as seen from Earth. Seven parameters (the four intrinsic longitudes of the spot, the single latitude of the spot, and the obliquity and rotation between the two coordinate systems) were fitted to eight measured quantities (the projected latitude and longitude of the spot during each of the four transits), and uncertainties determined using a simple Monte Carlo method. We found the true orbital obliquity to be $\psi = 20^\circ \pm 50^\circ$ and conclude that the available measurements are insufficient to put a strong constraint on this quantity. The spot positions are consistent with an aligned orbit, and are not consistent with a pole-on configuration. Better results could be obtained if star-spot positions could be measured at a wider range of longitudes and with increased precision in the measured latitudes, a situation which likely requires space-based data.

$\psi$ can be also constrained by estimating the stellar spin inclination angle, $i_{\star}$, knowing the rotational period of the parent star, that is
\begin{equation}
\sin{i_{\star}}=P_{\rm rot} \frac{(v \sin{i_{\star}})}{2 \pi R_{\star}}=1.07 \pm 0.40 \;,
\end{equation}
where we used $(v \sin{i_{\star}})=3.6 \pm 0.9$\,km\,s$^{-1}$ \citep{hebrard:2013}. Since values $>1$ are unphysical, $i_{\star}$ must have a value between 1 and 0.73, i.e. between $90^{\circ}$ and $47^{\circ}$. Then, using equation (7) from \citet{winn:2007}, 
\begin{equation}
\cos{\psi}=\cos{i_{\star}}\cos{i}+\sin{i_{\star}}\sin{i}\cos{\lambda}, %
\label{Eq:2}
\end{equation}
we estimated that ${\psi}$ is comprised between $6^{\circ}$ and $43^{\circ}$, which is in agreement with our previous measurement and also excludes that WASP-52\,A is in a pole-on configuration.

\section{Physical parameters of the WASP-52 planetary system}
\label{sec_4}

We used the {\it Homogeneous Studies} (HSTEP) approach (see \citealp{southworth:2012} and references therein) for revising the main physical properties of the WASP-52 planetary system. First of all, we established an input set of parameters, composed of:
\begin{itemize}
\item $r_{\rm A}+r_{\rm b}$, $k$, $i$, $P$, which were measured from the photometric light curves (this work; see Sect.\,\ref{sec_3});
\item $T_{\mathrm{eff}}=5000 \pm 100$ and [Fe/H]$=0.03 \pm 0.12$ measured from the spectroscopic analysis \citep{hebrard:2013};
\item the velocity amplitude of the star, $K_{\rm A}=84.3 \pm 3$\,m\,s$^{-1}$, measured from the radial velocities \citep{hebrard:2013}.
\end{itemize}

The orbital eccentricity was fixed to zero. We started the analysis by estimating the radial--velocity amplitude of the planet, $K_{\rm b}$, and determining an initial set of the physical parameters of the system, in particular the stellar mass. Using various tables of stellar parameters predicted by different theoretical models (i.e. Claret \citealp{claret:2004}; Y$^2$ \citealp{demarque:2004}; DSEP \citealp{dotter:2008}; VRSS \citealp{vandenberg:2006}; BaSTI \citealp{pietrinferni:2004}), we then interpolated to find the stellar radius and $T_{\mathrm{eff}}$ for our provisional mass and the observed [Fe/H], over all possible ages for the star.
After that, we adjusted $K_{\rm b}$ in an iterative way, with the aim to maximise the agreement between the measured values of $R_{\mathrm{A}}/a$ and $T_{\mathrm{eff}}$ and those predicted by one of the sets of theoretical models. We ended with a set of five values for each output quantity and we considered the unweighted mean of these as the final value. They are shown in Table\,\ref{tab:finalparameters}. Finally, we assigned two uncertainties for each of the final values: a systematic error based on the level of agreement among the values obtained using different theoretical models, and a statistical error related to the propagation of the uncertainties of the input parameters. Two parameters, $\rho_{\rm A}$ and $g_{\rm b}$, have no systematic error, as they can be directly estimated from observable quantities.

\subsection{Gyrochronological--isochronal age discrepancy}
\begin{figure*}%
{{\includegraphics[width=16.0cm]{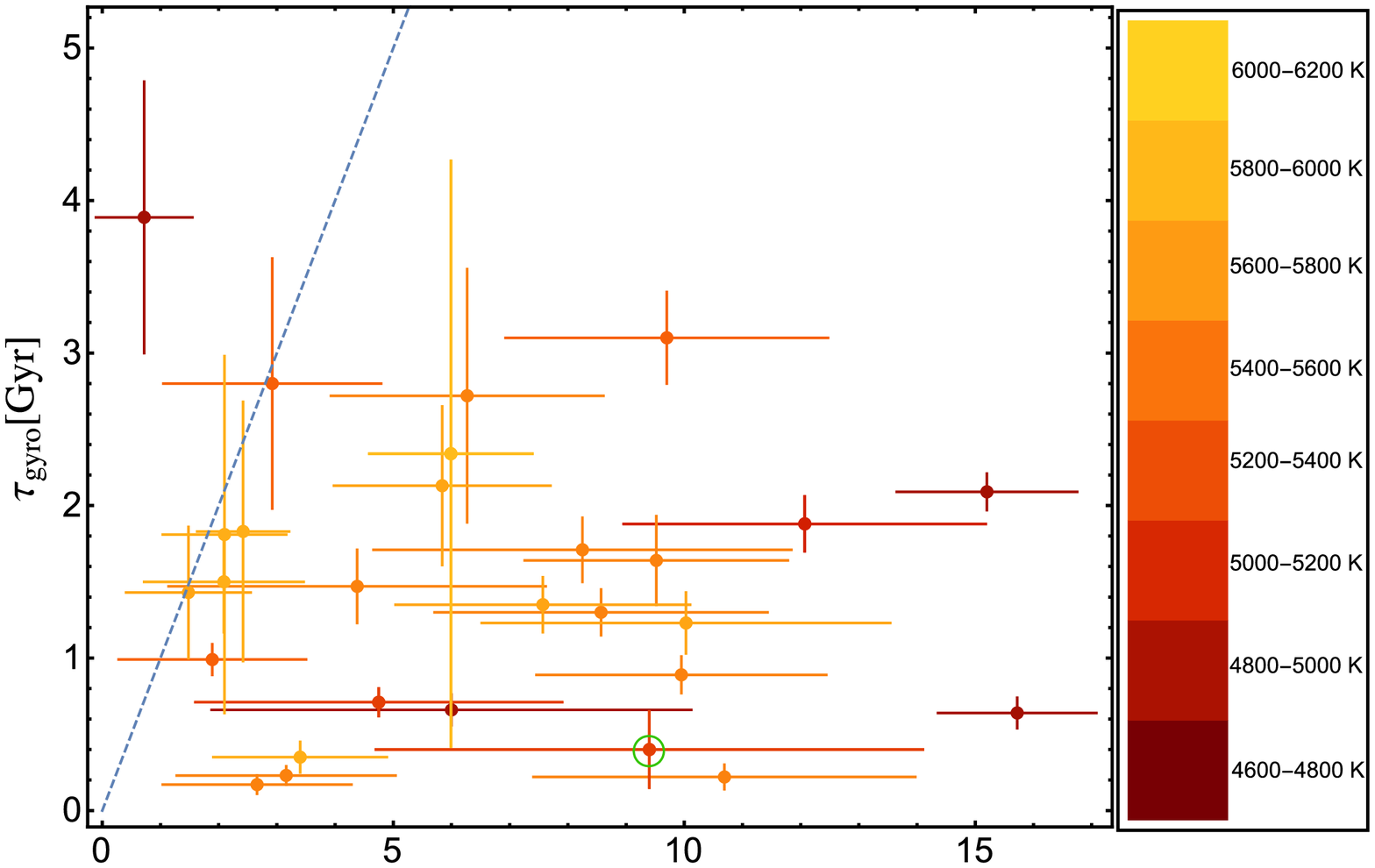} }}%
\qquad
{{\includegraphics[width=16.0cm]{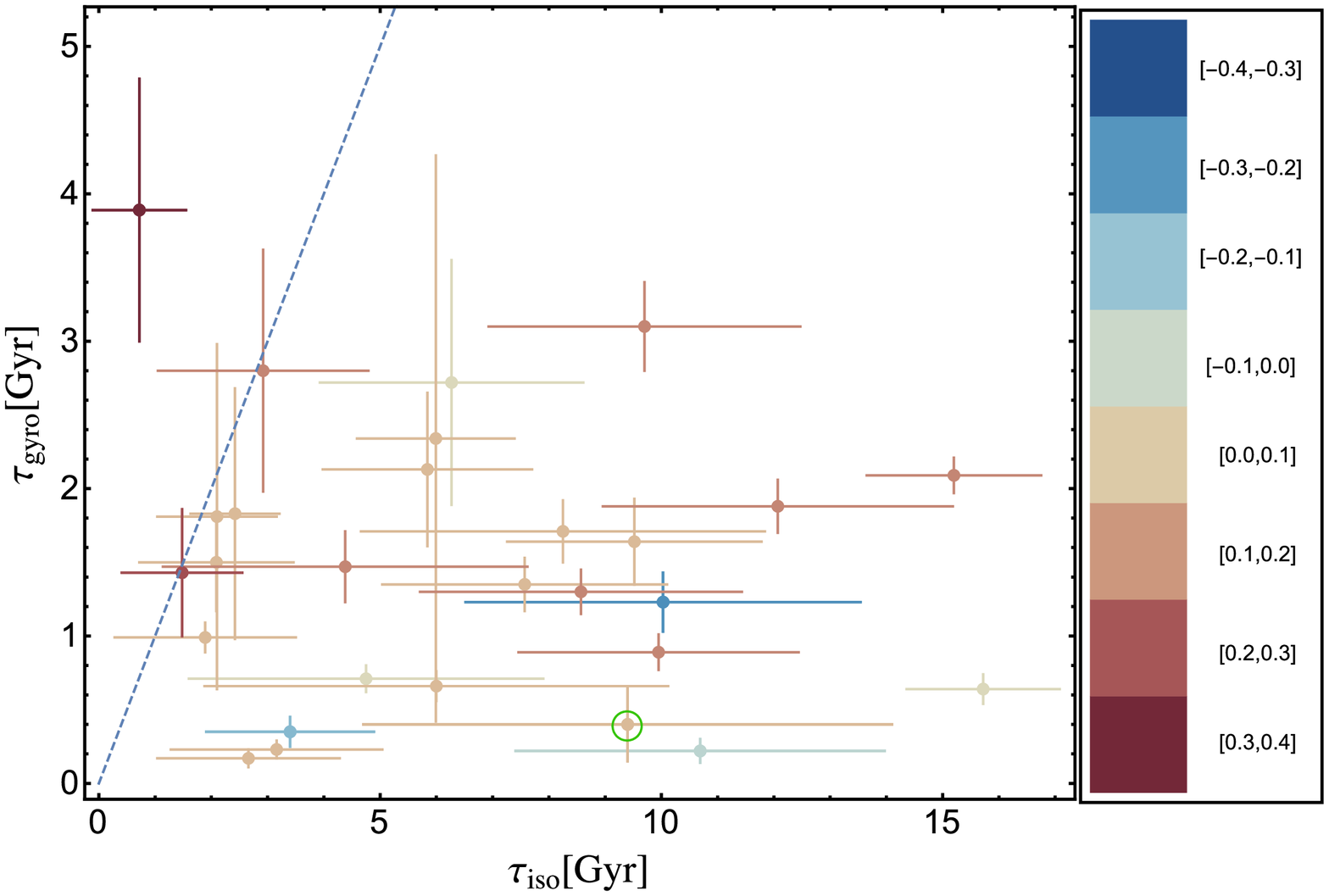} }}%
\caption{Plot of gyrochronological age estimates ($\tau_{\rm gyro}$) versus isochronal age
estimates ($\tau_{\rm iso}$) for hot-Jupiter parent stars with measured rotation periods
(list compiled by  \citealt{maxted:2015}). The position of WASP-52 (this work) is highlighted
with a green circle. The dashed line represents points for which $\tau_{\rm gyro}=\tau_{\rm iso}$.
Top panel: points are coloured according to the temperature of the corresponding star.
Bottom panel: points are coloured according to the metallicity of the corresponding star.}%
\label{fig:age}%
\end{figure*}

Almost all of our results, reported in Table\,\ref{tab:finalparameters}, are in good agreement with those found by \citet{hebrard:2013}. We found a slightly lower value for the stellar mass, but the two estimates are compatible within their uncertainties. However, the stellar age estimates are completely different. This fact is not a surprise as the two values were obtained using different procedures: our estimate is based on theoretical models (isochronal age, $\tau_{\rm iso}$), while that of \citet{hebrard:2013} comes from the star's rotation period (gyrochronological age, $\tau_{\rm gyro}$). It is well known that for many planetary systems, composed by a main-sequence star and a hot Jupiter, the gyrochronological age is significantly lower than the isochronal age (e.g.\ \citealt{pont:2009,lanza:2010,brown:2014}). This is clearly shown in Fig.\,\ref{fig:age}, where the two estimates for WASP-52 are plotted versus each other, together with those taken from the sample analysed by \citet{maxted:2015}. More than intrinsic stellar characteristics (like temperature or metallicity), the discrepancy is reasonably attributable to the star--planet tidal interactions in some cases; tides are actually able to transfer angular momentum from the orbit of a hot Jupiter to the rotation of the parent star, which are thus `spun up' and forced to rotate faster.

However, stellar activity can also play an important role for explaining this phenomenology. As noted by \citet{maxted:2015}, the $\tau_{\rm iso}-\tau_{\rm gyro}$ discrepancy is particularly evident for K-type stars, like WASP-52, and some G stars. The K stars suffer a `radius anomaly' (their size appears to be larger than predicted by standard stellar models; e.g. \citealt{popper:1997}), which is correlated with their rotation rate. The increase of the rotational velocity is, in turn, directly proportional to the amount of magnetic activity. A large number of star-spots can inhibit the efficiency of energy transport by convection, and affect the physical characteristics of the photosphere and the star's rotation rate.

That said, the age of WASP-52 estimated by gyrochronology \citep{hebrard:2013} could be underestimated due to the presence of the close giant planet and the magnetic activity of the star, which causes it to rotate faster. On the other hand, the star is clearly quite active and this suggests a young age, in contrast with the isochronal age that we have estimated. This situation is not easy to clarify and requires more sophisticated models.

\begin{figure*}
\centering
\includegraphics[width=16.0cm]{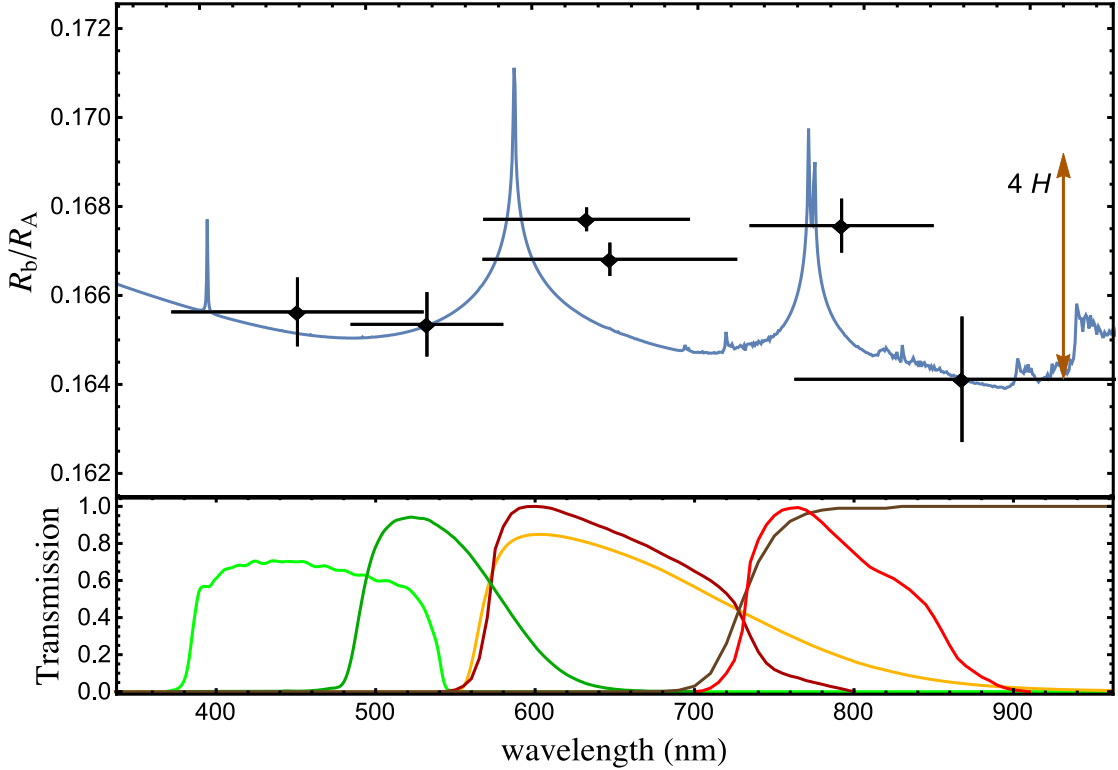}
\caption{Variation of the ratio of the planetary to stellar radii with wavelength. The black points correspond to the weighted mean $k$ values
obtained from the transit light curves analysed in this work. The vertical bars represent the relative uncertainties
and the horizontal bars show the FWHM transmission of the passbands. A synthetic spectrum for WASP-52\,b, obtained
with the \emph{petitCODE}, is shown as a continuous line and refers to a clear atmosphere. Offsets are applied to
the models to provide the best fit to our radius measurements. The atmospheres were computed for a planetary
metallicity the same as that of the parent star. The size of four atmospheric pressure scale heights ($4\,H$)
is shown on the right of the plot. Transmission curves of the filters are shown in the bottom panel.}
\label{fig:radiusvariation}
\end{figure*}

\begin{table*} \centering
\caption{Physical parameters of the planetary system WASP-52 derived in this work, compared with those from \citet{hebrard:2013}.
Where two error bars are given, the first refers to the statistical uncertainties, while the second to the systematic errors.
{\bf Notes}. $^{a}$Our estimate of the stellar age was derived from theoretical models, and that from the discovery paper was
obtained from the stellar rotation period. $^{b}$The Safronov number represents the ratio of the escape velocity to the orbital velocity of the planet and indicates the extent to which the planet scatters other bodies. $^{c}$Our measurement of the time of mid-transit is given in BJD, while that from \citet{hebrard:2013} is in HJD.
$^{d}$Our values for the spin-orbit angle were derived under the hypothesis
that the same star-spot complex was occulted by the planet in four close transit events.}
\label{tab:finalparameters}
\begin{tabular}{l c c c c} \hline
Quantity & Symbol & Unit & This work & \citet{hebrard:2013}\\
\hline  \\[-6pt]
\multicolumn{1}{l}{\textbf{Stellar parameters}} \\
Stellar mass   \dotfill                   & $M_{\rm A}$     & \Msun & $0.804 \pm 0.050 \pm 0.004$       & $0.87 \pm 0.03$     \\
Stellar radius  \dotfill                   & $R_{\rm A}$     & \Rsun & $0.786 \pm 0.016 \pm 0.001$       & $0.79 \pm 0.02$     \\
Stellar surface gravity  \dotfill     & $\log g_{\rm A}$& cgs   & $4.553 \pm 0.010 \pm 0.001$       & $4.5 \pm 0.1$     \\
Stellar density   \dotfill                & $\rho_{\rm A}$  & \psun & $1.653 \pm 0.020$                 & $1.76 \pm 0.08$     \\
Age$^{a}$  \dotfill                                  & $\tau$          & Gyr   & $9.4_{-4.3\,-1.4}^{+4.7\,+1.2}$   & $0.4^{+0.3}_{-0.2}$   \\
\hline \\[-6pt]%
\multicolumn{1}{l}{\textbf{Planetary parameters}} \\
Planetary mass  \dotfill               & $M_{\rm b}$     & \Mjup & $0.434 \pm 0.024 \pm 0.002$       & $0.46 \pm 0.02$     \\
Planetary radius  \dotfill              & $R_{\rm b}$     & \Rjup & $1.253 \pm 0.027 \pm 0.002$       & $1.27 \pm 0.03$     \\
Planetary surface gravity  \dotfill & $g_{\rm b}$     & \mss  & $6.85 \pm 0.26$                   & $6.46 \pm  0.45$       \\
Planetary density  \dotfill            & $\rho_{\rm b}$  & \pjup & $0.2061 \pm 0.0091 \pm 0.0004$       & $0.22 \pm 0.02$     \\[2pt]
Equilibrium temperature  \dotfill  & \Teq            & K     & $1315 \pm 26$                     & $1315 \pm 35$~~~         \\
Safronov number$^{b}$  \dotfill    & \safronov\      &       & $0.02273 \pm 0.00094 \pm 0.00004$    &      --                 \\
\hline \\[-6pt]%
\multicolumn{1}{l}{\textbf{Orbital parameters}} \\
Time of mid-transit         \dotfill & $T_{0}$        & BJD / HJD$^{c}$ & $2\,456\,862.79776 \pm 0.00016 $ & $2\,455\,793.68143 \pm 0.00009 $ \\ %
Period              \dotfill & $P_{\mathrm{orb}}$            & days    & $1.74978119 \pm 0.00000052 $ & $1.7497798 \pm 0.0000012 $ \\ %
Semi-major axis   \dotfill   & $a$             & au    & $0.02643 \pm 0.00055 \pm 0.00005$ & $0.0272 \pm 0.0003$ \\
Inclination         \dotfill & $i$            & degree  & $85.15   \pm 0.06   $ & $85.25   \pm 0.20   $ \\  [1pt] %
Projected spin-orbit angle$^{d}$ & $\lambda$  & degree   & $3.8 \pm 8.4$ & $24_{-9}^{+17}$   \\
True spin-orbit angle$^{d}$ & $\psi$  & degree   & $20 \pm 50$ & -- \\
\hline
\end{tabular}
\end{table*}

\begin{figure*}
\centering
\includegraphics[width=16.0cm]{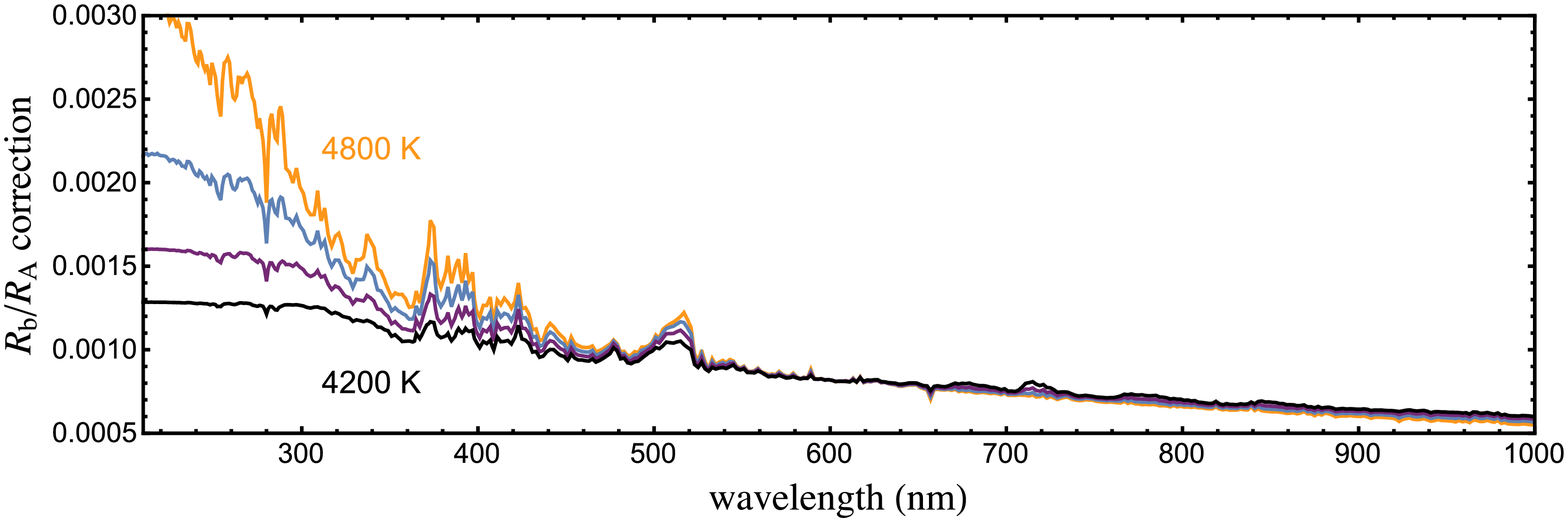}
\caption{The effect of unocculted star-spots on the transmission spectrum of WASP-52 b, considering a $1\%$ flux drop at 600\,nm. A stellar temperature of $T_{\rm eff}= 5000$\,K was adopted. The star-spot coverage was modelled using a grid of stellar atmospheric models at different temperature ranging from 4800K (yellow line) to 4200K (black line), in steps of 200\,K.}
\label{fig:spot_correction}
\end{figure*}

\section{Variation of the planetary radius with wavelength}
\label{sec_5}

The transmission spectrum of hot Jupiters is expected to show characteristic absorption features at particular wavelengths. In the visual region, some of them are due to sodium ($\sim 590$\,nm), potassium ($\sim 770$\,nm) and water vapour ($\sim 950$\,nm). However, the variety of hot-Jupiter transmission spectra suggests a great deal of variation in chemistry and atmospheric dynamics, and some of them can be dominated by Mie or Rayleigh scattering (e.g.\ \citealt{sing:2016}). Using light curves taken through different passbands, we made an attempt to reconstruct the transmission spectrum of WASP-52\,b. Following the approach used in previous studies (e.g.\ \citealp{southworth:2015,mancini:2016}), we refitted the light curves to estimate the ratio of the radii, $k$, whilst fixing the other photometric parameters to their best values (Tables \ref{Table:fits} and \ref{tab:finalparameters}). The corresponding errorbars were calculated by performing 10\,000 Monte Carlo simulations. In this way, we obtained new values of $k$, whose errorbars do not include common sources of uncertainty. These are shown in Fig.\,\ref{fig:radiusvariation} and compared with a synthetic spectrum, which is based on a self-consistent modelling of one-dimensional atmospheric structures, obtained with a new version of the {\it petitCODE} \citep{molliere:2015,mancini:2016b}. The theoretical model represents the case of a clear atmosphere for WASP-52\,b, without opacities caused by strong absorbers such as gaseous titanium oxide.

The observations show a flat transmission spectrum to within the experimental uncertainties; the maximum planetary radius variation is between the LNIR (Baronnies telescope) and the Bessel-$R$ (Danish telescope) bands, but the detection is 2.4 pressure scale heights\footnote{The pressure scale height is defined as $H = \frac{k_{\rm B}T_{\rm eq}}{\mu_{\mathrm{m}}\,g_{\mathrm{p}}}$, where $k_{\rm B}$ is the Boltzmann's constant, $\mu_{\mathrm{m}}$ the mean molecular weight, $T_{\rm eq}$ the planetary equilibrium temperature, $g_{\rm p}$ the planetary surface gravity.} with a confidence level of just $2.4 \sigma$. We stress that this result is not significant and is based on light curves taken with different instruments and at different times.
Moreover, unocculted star-spots can cause variations of the transit depth, which are dependent on wavelength and the amount of stellar activity at particular cycles.
These variations can be particularly stronger at bluer wavelengths. We estimated the effect of unocculted star-spots on the transmission spectrum of
WASP-52\,b using the methodology described by \citet{sing:2011}. The correction to the transit depth for unocculted star-spots is shown in Fig.\,\ref{fig:spot_correction} for
different star-spot temperatures, assuming a total dimming of $1\%$ at a reference wavelength of 600\,nm \citep{sing:2011}. This effect is very small, and is well inside the observational uncertainties.

\section{Summary and discussion}
\label{sec_6}
\begin{figure}
\centering
\includegraphics[width=\columnwidth]{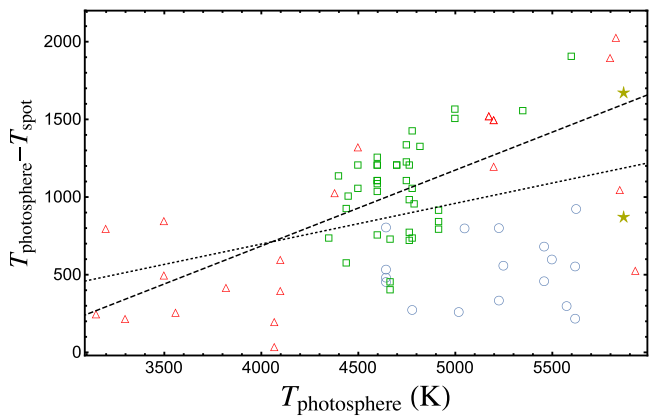}
\caption{Spot temperature contrasts from transiting planetary systems, blue circles, and published values (taken from \citealt{andersen:2015}); red triangles are related to main-sequence dwarf stars and green squares to main-sequence giant stars. Contrast values for the Sun are indicated by yellow stars and represent the umbral
(higher) and penumbral (lower) temperature contrasts \citep{berdyugina:2005}. Dashed line is the linear fit to all the points after excluding those coming from the transits (Pearson linear correlation coefficient $r=0.67$), while dotted line is the linear fit to all the points (Pearson linear correlation coefficient $r=0.36$).}
\label{fig:spot_contrast_2}
\end{figure}

In this work we reported photometric observations of eight transit events of WASP-52\,b, performed using four different medium-class telescopes, located on both of Earth's hemispheres, through different optical passbands. All of the transits were observed using the \emph{defocussing} technique, achieving a photometric precision of $0.51-1.79$\,mmag per observation. Two transits were simultaneously monitored with two different telescopes, once at the same observatory, and once in different countries. In the former case, a multi-band imaging camera was used. In total, we have presented 13 new light curves. Light-curve anomalies have been clearly noted in five transits. Considering the spectral class of the parent star, these anomalies are reasonably explained by star-spot occultation events caused by the planet during its transits. The light curves and the anomalies were modelled and their main parameters determined. Our principal results are as follows.
\begin{itemize}
\item We have used these new light curves, plus data taken from the literature and from the ETD archive, to refine the orbital ephemeris and the physical parameters of the WASP-52 planetary system. Our results are shown in Table\,\ref{tab:finalparameters} and are in a good agreement with those measured by \citet{hebrard:2013} (the star-spot contamination on their data was not strong enough to influence their estimation of the planetary-system parameters). The only exception is the age of the system, which is very discordant. Such a discrepancy can be explained because of the different methods used in the two analyses: we reported an age based on theoretical models, while \citet{hebrard:2013} used gyrochronology. The isochronal age is not compatible with the activity of the star, but the gyrochronological age could be severely underestimated, as the presence of the close-in hot Jupiter and, again, stellar activity were not taken into account \citep{maxted:2015}. This case is emblematic of the limits of the techniques currently used to estimate stellar ages. \\

%
\item We carefully characterised the star-spots detected in various transits. Our best-fitting models yield measurements of their positions on the stellar disc, and their size and contrast. From these we extracted their temperature and speculate about the alignment between the planet's orbital axis and the stellar spin.

\begin{itemize}
\item We estimated the star-spot temperature contrast for WASP-52 and compare it with those derived from ($i$) the light curves of other transiting planetary systems and ($ii$) via other techniques. Joining the various data sets in a global picture, the dependence of the starpot temperature contrast with the spectral class is not anymore evident as in the case in which the data from transits were not considered, see Fig.\,\ref{fig:spot_contrast_2}. \\ [-8pt]
\item We found a sky-projected orbital obliquity of $\lambda = 3^{\circ}.8 \pm 8^{\circ}.4$, which is consistent with and more precise than the value found from the Rossiter-McLaughlin effect \citep{hebrard:2013}. Using the four positions measured for the same star-spot, we were able to place a weak constraint on the true orbital obliquity: $\psi = 20^\circ \pm 50^\circ$. To our knowledge, this is the first measurement of $\psi$ based on only star-spot crossings.
\end{itemize}
\item Since our transits were recorded through different filters at optical wavelengths, we attempted to reconstruct an optical transmission spectrum of the planet. We found a small variation, $2.4\,H$, of the planet's radius, but at a low significance. We conclude that the transmission spectrum of WASP-52 is flat to within the experimental errors. However, more precise and simultaneous multi-band observations are suggested to robustly confirm our finding.

\end{itemize}

\section*{Acknowledgements}

This paper is based on observations collected with ($i$) the Zeiss 1.23\,m telescope at the Centro Astron\'{o}mico Hispano Alem\'{a}n (CAHA) at Calar Alto, Spain; ($ii$) the Danish 1.54\,m telescope at the ESO Observatory in La Silla, Chile; ($iii$) the Cassini 1.52\,m telescope at the Astronomical Observatory of Bologna in Loiano, Italy; ($iv$) the MPG 2.2\,m telescope located at the ESO Observatory in La Silla, Chile. Operations at the Calar Alto telescopes are jointly performed by the Max Planck Institute for Astronomy (MPIA) and the Instituto de Astrof\'{i}sica de Andaluc\'{i}a (CSIC). Operation of the Danish 1.54\,m telescope is financed by a grant to UGJ from the Danish Natural Science Research Council (FNU). Operation of the MPG 2.2\,m telescope is jointly performed by the Max Planck Gesellschaft and the European Southern Observatory. GROND was built by the high-energy group of MPE in collaboration with the LSW Tautenburg and ESO, and is operated as a PI-instrument at the MPG 2.2\,m telescope. OW and J. Surdej acknowledge support from the Communaut\'{e} fran\c{c}aise de Belgique -- Actions de recherche concert\'{e}es -- Acad\'{e}mie Wallonie-Europe. TCH acknowledges financial support from the Korea Research Council for Fundamental Science and Technology (KRCF) through the Young Research Scientist Fellowship Programme and is supported by the KASI research grant 2014-1-400-06 and 2016-1-832-01.
The reduced light curves presented in this work will be made available at the CDS (http://cdsweb.u-strasbg.fr/).
We thank the anonymous referee for their useful criticisms and suggestions that helped us to improve the quality of this paper.
We thank Matthias Mallonn for his useful comments.
The following internet-based resources were used in research for this paper: the ESO Digitized Sky Survey; the NASA Astrophysics Data System; the SIMBAD data base operated at CDS, Strasbourg, France; the arXiv scientific paper preprint service operated by the Cornell University.



\bsp 
\label{lastpage}
\end{document}